%% file: jcapexample.tex
\documentclass[a4paper,11pt]{article}
\pdfoutput=1 

\usepackage{jcappub} 

\usepackage[T1]{fontenc} 
\usepackage{enumitem}

\input{local-macro}

\title{Consistency Relations in Effective Field Theory}
\author{Dipak Munshi$^{1}$, Donough Regan$^{2}$}
\affiliation{Astronomy Centre, School of Mathematical and Physical Sciences,\\ University of Sussex, Brighton BN1 9QH, U.K.}
\emailAdd{$^1$D.Munshi@sussex.ac.uk, $^2$D.Regan@sussex.ac.uk}

\abstract{The consistency relations in large scale structure relate the lower-order 
correlation functions with their higher-order counterparts. They are direct
outcome of the underlying symmetries of a dynamical system and can be tested using data from future surveys such as Euclid. Using techniques from 
standard perturbation theory (SPT), previous studies of consistency relation have concentrated on continuity-momentum (Euler)-Poisson system
of an ideal fluid. We investigate the consistency relations in effective field theory (EFT) which adjusts the SPT predictions
to account for the departure from the ideal fluid description on small scales. We provide detailed results for the 
3D density contrast $\delta$ as well as the {\em scaled} divergence of velocity $\bar\theta$. Assuming a $\Lambda$CDM background cosmology, we find the correction to 
SPT results becomes important at $k \gtrsim 0.05 \rm h/Mpc$ and that
the suppression from EFT to SPT results that scales as square of the wave number $k$, can reach $40\%$ of the total at $k \approx 0.25\rm h/Mpc$ at $z=0$. We have also investigated whether effective field theory corrections to models of primordial non-Gaussianity can alter the squeezed limit behaviour, finding the results to be rather insensitive to these counterterms. In addition, we present the EFT corrections to the squeezed limit of the bispectrum in redshift space which may be of interest for tests of theories of modified gravity.

\par\vspace{50mm}\mbox{}\hfill
\raisebox{-0.5\height}{\includegraphics[scale=0.2]{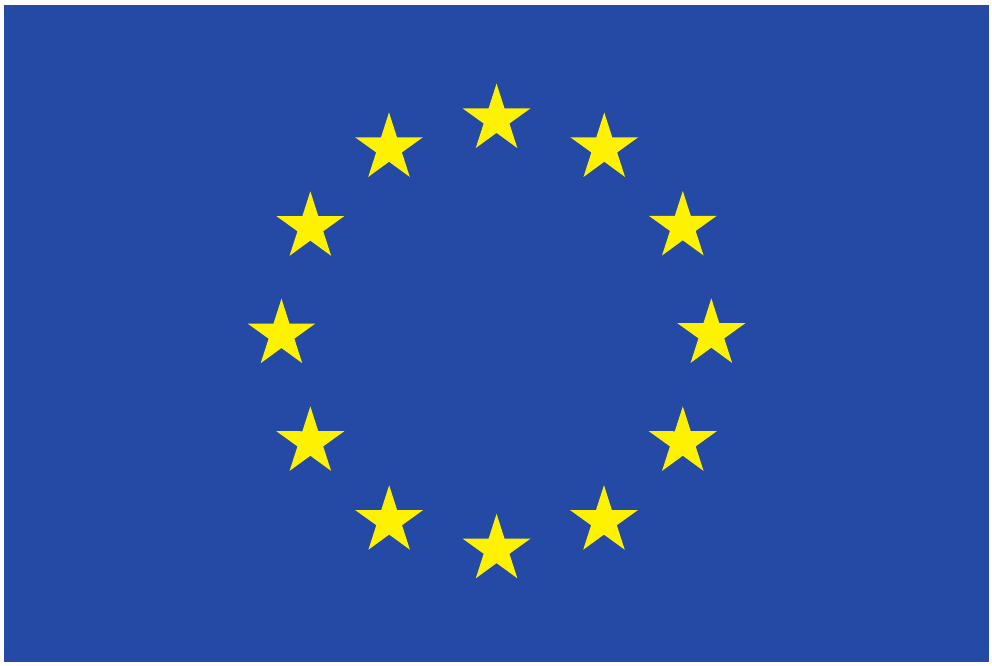}}
\;{\vrule width 1pt}\;
\raisebox{-0.5\height}{\includegraphics[scale=0.1]{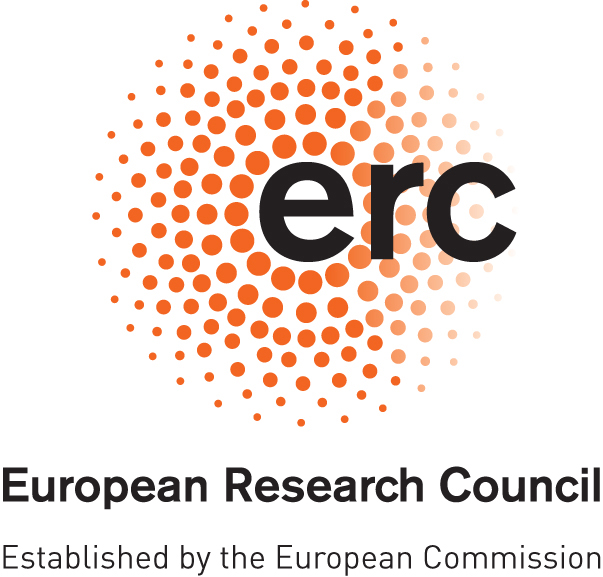}}
\quad
\raisebox{-0.5\height}{\includegraphics[scale=0.5]{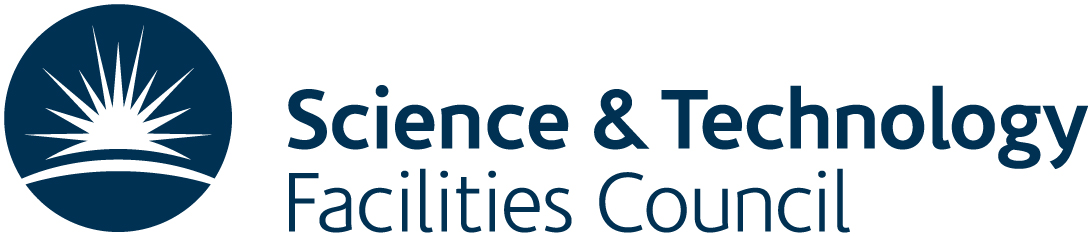}}}


\begin{document}
\maketitle
\flushbottom

\section{Introduction}
After the completion of the 
Planck\footnote{Planck: \href{https://www.cosmos.esa.int/web/planck}{\tt https://www.cosmos.esa.int/web/planck}} mission \citep{Planck}, large scale surveys are expected to play a more dominant role
in answering many of the puzzling questions of modern day cosmology. Several ongoing surveys are already mapping the
large scale matter distribution with ever increasing precision while others are being 
planned (e.g. BOSS\footnote{Baryon Oscillator Spectroscopic Survey: 
\href{http://www.sdss3.org/surveys/boss.php}{\tt http://www.sdss3.org/surveys/boss.php}}
\citep{EW},
WiggleZ\footnote{WiggleZ Survey : \href{http://wigglez.swin.edu.au/}{\tt http://wigglez.swin.edu.au/}}
\citep{DJA},
LSST\footnote{The Large Synoptic Survey Telescope : \href{https://www.lsst.org/}{\tt http://https:/www.lsst.org/}},
DES\footnote{Dark Energy Survey: \href{http://www.darkenergysurvey.org/}{\tt http://www.darkenergysurvey.org/}}
\citep{DES},
EUCLID\footnote{EUCLID: \href{http://www.euclid-ec.org/}{\tt http://www.euclid-ec.org/}}
\citep{LAA}).

%
%
Cosmic Microwave Background (CMB) experiments primarily 
probe the high redshift Universe where the growth of inhomegeneities are in the linear regime and are
relatively easy to understand. By contrast, the growth of large scale structure at lower redshifts
involves nonlinear physics of gravitational clustering characterized by 
the coupling of small and large scale modes. This gravitational clustering
is typically studied using a perturbative framework, known commonly as standard perturbation theory (SPT), and its extensions 
\citep{review,resum,Takahashi, Shoji_Komatsu, Jeong_Komatsu, Crocce1, Crocce2, Matsubara, McDonald, Taruya, IzumiSoda,equal,Sabino1,Sabino2,Nishi,Peebles, Carlson,Pietroni, Zaldarriaga,Porciani} - however, these
results are valid only in the quasilinear regime and tend to break down at smaller length scales.
Renormalized perturbation theory or (RPT) was introduced to improve the performance of SPT \citep{Crocce1,Bernard_Crocce},
by dividing the nonlinear corrections into (a) a mode-coupling effect and (b) an appropriate normalization of the propagator. The RPT approach, which attempts to solve the full non-linear equations for a pressureless ideal fluid as exactly as possible, is known to significantly improve the performance of the SPT, but does not address the effect of backreaction due to ultraviolet (UV) physics.

Most perturbative approaches have limited range of applicability, and in the highly non-linear regime, sophisticated numerical algorithms, such as {\tt{GADGET}}\footnote{GADGET: \href{https://www.mpa.mpa-garching.mpg.de/gadget/}{\tt https://www.mpa.mpa-garching.mpg.de/gadget/}} \citep{Springel:2005mi}, exist to evolve a self-gravitating 
system. These simulations, however, can be highly expensive for detailed exploration of the entire cosmological
as well as astrophysical parameters. Different phenomenological approaches, such as the halo model prescription \cite{halo}, are also 
used regularly for making quantitative predictions \citep{halo}. An alternative treatment -- known as the effective field theory of large scale structure -- to \textit{reduce} the necessity of expensive simulations while retaining analytic control was introduced recently in \citep{Baumann,EFT1,Vlah,Porto} (also see \citep{Pietroni}  for a similar approach). 

%
%
The effective field theory (EFT) is based on exploitation of the symmetries of a system to incorporate small-scale effects
that are beyond the analytical treatment. The large scale modes,
which can be modelled analytically, are dealt with using standard perturbation
theory (SPT). The EFT provides a self-consistent framework to
handle the coupling of large- and small-scale modes through the inclusion of a stress tensor, thus
equivalently forcing us to deal with an non-ideal fluid. 
The EFT approach remedies many of the fundamental issues faced by 
the SPT. Indeed, SPT is plagued by UV-divergent integrals
resulting from a high scale cut-off that appears in loop level corrections \citep{PajerZal}.
It also deals with the breaking down of the assumption of a pressure-less ideal fluid on small scales. In EFT, these divergences are normalized by introducing ``counter-terms'' that take the form of a non-ideal fluid with the inclusion of
non-zero pressure, viscosity and thermal conductivity terms. These unknown parameters 
describing a non-ideal fluid, however, can not be computed within the EFT framework, but can be estimated using
numerical simulations on small scales.

There is an extensive literature by many
different authors in this area as it has emerged as a very active area of research. Our approach is
closest to the one described in \citep{EFT1,EFT2}, with our discussion of the EFT bispectrum largely following the description in \citep{bispec}, where it was shown to be be accurate  up to $k\approx 0.22\,h\, \rm Mpc^{-1}$
for $z=0$ (a factor of two better than achievable with one-loop SPT). Of particular relevance to the study performed in this paper are results regarding the statistics of velocity in EFT presented in \citep{velocity}, primordial non-Gaussianity in \citep{primordial,Yvette} and on redshift space distortions in \citep{redshift,Perko:2016puo,delaBella:2017qjy}.

%
%
Consistency relations encode correlations between large-scale linear modes and
small-scale non-linear modes.
The advantage of these relations
stems from the fact that they are valid despite our poor understanding
of the non-linear gravitational clustering or the complicated astrophysics
of star formation and supernovae feedback. These relations are
kinematic and simply enforce the equivalence principle so that, at leading order where large scale forces are assumed to be constant, the small structures are
transported without distortion by the large-scale fluctuations. The consistency relations can, therefore, act as an important diagnostics for detection any departure 
from General Relativity \citep{Munshi2}. Consistency relations for the CMB secondaries were investigated in \citep{Munshi1,const5}, while
the consistency conditions of large scale structure observables, studied in \citep{equal,Peloso,Creminelli1, Creminelli2,Valageas,const2}, were derived recently in the context of redshift space distortions in \citep{const1}, and
in the presence of primordial non-Gaussianity in \citep{const4}. The density-velocity consistency relations were 
derived in \citep{const3}.

This article is organized as follows: in Sec.~\ref{sec:eftoflss} we summarize the effective field theory formalism, reviewing the results for the density and velocity-divergence bispectra. In Sec.~\ref{sec:PNG} we generalise the discussion to primordial non-Gaussianity, which introduces an extra degree of freedom into the construction of the effective stress-energy tensor, which encapsulates the UV physics. This is followed in Sec.~\ref{sec:RSD} by a discussion of the bispectrum in redshift space where the distorting effect of peculiar velocities along the radial direction must be accounted for. Sec.~\ref{sec:consistencyRels} presents the key new results of this paper. We begin by elucidating the relationship between the consistency relations and the angle-averaged squeezed bispectrum, and in turn the integrated bispectrum. We then compute the squeezed bispectra for each of the previously described cases, to evaluate the importance of accounting for the impact of UV physics on each of the observables. Finally, in Sec.~\ref{sec:conclu} we present our conclusions, and proposals for further applications of this line of investigation.

%
\section{Effective Field Theory of Large Scale Structure}
\label{sec:eftoflss}
The standard perturbation theory (SPT) framework is given by the non-relativistic limit of a perfect fluid of pressureless (cold) dark matter in the Newtonian gauge. On large scales this prescription agrees well with N-body simulations. However, on somewhat smaller scales more accurate predictions necessitate integrals over convolutions of first order solutions. These convolutions couple large and small scales, the latter of which are impacted by ultraviolet (UV) physics which violate the perfect fluid approximation. The effective field theory framework developed in \citep{Baumann,EFT1,2loop,Porto,resum,redshift} can be regarded as a renormalisation of the density and velocity perturbations by introducing counterterms which aggregate the effect of the UV physics \citep{delaBella:2017qjy}. Equivalently one can regard the effect of the small scale physics as generating terms like viscosity and stochasticity in an effective stress-energy tensor. The equations of motion in comoving time $\tau=\int dt/a(t)$ takes the following form for the density, $\delta$, and velocity divergence $\theta\equiv {\bf \nabla. v}$:
\ben
\label{eq:cont}
&&{\partial \over \partial \tau} \delta + \theta = S_{\alpha};\\
&& {\partial\theta \over \partial {\tau}} + {\cal H}\theta + {3\over 2}\Omega_{\rm M}{\cal H}^2\delta = S_\beta+\tau_\theta\,,
\label{eq:euler}
\een
where the convolutions $S_{\alpha}(\bk,\tau)$ and $S_\beta(\bk,\tau)$ denote
\bes
\ben
\label{eq:alpha}
&& S_{\alpha}(\bk,\tau)\equiv \alpha[\bp\star(\bk-\bp)] \equiv -\int \alpha(\bp,\bk-\bp)\theta^{}(\bp,\tau)\delta^{}(\bk-\bp,\tau)\,,\\
&&S_{\beta}(\bk,\tau)\equiv \beta[\bp\star(\bk-\bp)]  \equiv -\int \beta(\bp,\bk-\bp)\theta^{}(\bp,\tau)\theta^{}(\bk-\bp,\tau)\,.
\label{eq:beta}
\een
\ees
The kernels $\alpha(\bk_1,\bk_2)$ and $\beta(\bk_1,\bk_2)$ are, in turn, given by:
\ben
\alpha(\bk_1,\bk_2) \equiv {\bk_{12}}\cdot{\bk_1 \over k_1^2} \quad \text{ and}\quad \beta(\bk_1,\bk_2) \equiv {1\over 2} \bk^2_{12}{\bk_1\cdot\bk_2 \over k_1 k_2}; \quad
\bk_{1\cdots p}= \bk_1+\cdots +\bk_p.
\een
To these one also appends the Poisson equation, relating the density to the gravitational potential via
\ben\label{eq:Poisson}
&& \triangle \phi = {3\over 2}{\cal H}^2 \Omega_M \delta\,,
\een
where ${\cal H}\equiv \partial \ln a/\partial \tau$, and $\Omega_M$ is the dark matter abundance, as usual. These equations are identical, in the case of negligible vorticity, to the SPT equations apart from the presence of the divergence of the effective-stress-energy tensor term, $\tau_\theta$, which incorporates the UV physics. The form of this term may be deduced on symmetry grounds. To second order, where stochastic terms can be neglected one writes (in the absence of primordial non-Gaussianities which shall be dealt with in Sec.~\ref{sec:PNG}) \citep{bispec}
\ben
&& \tau_{\theta} = -d^2\triangle\delta -e_1\triangle(\delta^2) -e_2\,\triangle(s^2)
-e_3\partial_i(s^{ij}\partial_j\delta)\,,  
\label{eq:ansatz_noPNG}
\een
where $d^2$ can be thought of as an aggregated term for the speed of sound or viscosity, while $s_{ij}\equiv(\partial_i \partial_j - \triangle \delta_{ij}/3)\Phi$ is the tidal term, for $\Phi\equiv 2\phi/( {3}{\cal H}^2 \Omega_M)$, with $s^2=s_{ij}s^{ji}$.

\para{Green's function solutions} Equations~\eqref{eq:cont} and~\eqref{eq:euler} may be solved perturbatively using a Green's function approach such that
\bes
\ben
&&\delta(\bk,a) = D_+(a)\delta^{(1)}(\bk) + \int^a_0 {\cal G}_{\delta}(a,a^{\prime})\left [S_\beta +\tau_\theta -{\cal H}\partial_{a^{\prime}}(a^{\prime} S_\alpha)\right] da^{\prime}\,,\\
&&\theta(\bk,a) = -{\cal H} f D_+(a)\delta^{(1)}(\bk) + \int^a_0 {\cal G}_{\theta}(a,a^{\prime})\left [\partial_{a^{\prime}}(a^{\prime} S_\beta + a^{\prime}\tau_\theta) - \frac{3}{2}\Omega_M {\cal H}S_\alpha\right] da^{\prime}\,,
\label{eq:gen}
\een
\ees
where $\delta^{(1)}(\bk)$ denotes the initial density perturbation, $D_+$ represents the growth factor, while $f\equiv d\ln D_+/d\ln a$ is dubbed the growth rate. The Green's functions  ${\cal G}_\delta$ and ${\cal G}_\theta$ are given by:
\bes
\ben
&& {\cal G}_{\delta}(a,a^{\prime}) = {5\over 2}\Theta(a-a^{\prime})\Omega_0 {\cal H}_0^2 {{\cal H}\over a}\left \{ {D_{-}(a^{\prime}) \over D_{-}(a)}-{D_+(a^{\prime})\over D_+(a)} \right \}; \\
&&  {\cal G}_{\theta}(a,a^{\prime}) = -{\cal H}{\cal G}_{\delta}(a,a^{\prime}).
\een
\ees
Here $D_-$ represents the decaying modes and $\Theta$ the unit step function, and the `$0$' subscript is used to denote quantities evaluated at the present time. 

Iteratively solving equations~\eqref{eq:gen} allows one to obtain the general solution to arbitrary order. One may separate the SPT solution (with source $\tau_\theta$ set to zero) and write,
\ben\label{eq:DeltaThetaExpansion}
 \delta(\bk,a)=\sum^{\infty}_{n=1}[\delta^{(n)}(\bk,a) + \delta_{c}^{(n)}(\bk,a)]\,,\,\,\,{\text{and}}\,\,\, \theta(\bk,a)=-{\cal H} f\sum^{\infty}_{n=1}[\bar{\theta}^{(n)}(\bk,a) + \bar{\theta}_{c}^{(n)}(\bk,a)]\,,
\een
with $\delta^{(n)}$ ($\delta_{c}^{(n)}$) and $\bar{\theta}^{(n)}$ ($\bar{\theta}_{c}^{(n)}$) denoting the SPT (EFT) contributions. Substituting the first order SPT solution $\delta^{(1)}$ into $\tau_\theta$ one utilises \eqref{eq:gen} with the other sources set to zero to obtain the leading order EFT solution. The EFT solutions at second order are similarly obtained using the Green's function method with source given by
\begin{enumerate}[label=(\alph*)]
\item $\tau_\theta$ only at second order, as obtained using Eq.~\eqref{eq:ansatz_noPNG} using the first and second order SPT solutions for $\delta$ and $\theta$.
\item the entire source term of Eq.~\eqref{eq:gen}, but with $\tau_\theta=0$, evaluated using the first order EFT solution (with the first order SPT solution). Such solutions will be called the `$\alpha\beta$' terms for reasons that will be apparent later. 
\end{enumerate}
\para{Einstein-de Sitter (EdS) solutions} In the Einstein-de Sitter limit, $\Omega_M=1$ (and $f=1$), the time dependence of the SPT terms factorise simply ($\propto a^n$ for $\delta^{(n)}$) and may be written as
\bes\label{eq:EdS_solution}
\ben 
&& \delta^{(n)}(\bk,a) \approx D_+^n(a)\int \frac{d^3\bk_1}{(2\pi)^3}\cdots \frac{d^3\bk_n}{(2\pi)^3}[\delta_D]_n F_n(\bk_1,\cdots,\bk_n)\delta^{(1)}(\bk_1)\cdots\delta^{(1)}(\bk_n)\,,\\
&&  \bar{\theta}^{(n)}(\bk,a) \approx D_+^n(a)\int \frac{d^3\bk_1}{(2\pi)^3}\cdots \frac{d^3\bk_n}{(2\pi)^3}[\delta_D]_n G_n(\bk_1,\cdots,\bk_n)\delta^{(1)}(\bk_1)\cdots\delta^{(1)}(\bk_n)\,,
\een
\ees
where $[\delta_D]_n\equiv (2\pi)^3\delta_D(\bk-\bk_{1\cdots n})$, and iterative expressions for the kernels $F_n$ and $G_n$ are given in \citep{review}, with explicit formulae for $F_2$ and $G_2$ written
\bes
\ben
&& F_2(\bk_1,\bk_2)={5\over 7} +{1\over 2}\left({k_1\over k_2}+{k_2\over k_1}\right )\left({\bk_1\cdot\bk_2 \over k_1 k_2}\right ) +{2 \over 7}\left({\bk_1\cdot\bk_2 \over k_1 k_2}\right )^2\,,\\
&& G_2(\bk_1,\bk_2)={3\over 7} +{1\over 2}\left({k_1\over k_2}+{k_2\over k_1}\right )\left({\bk_1\cdot\bk_2 \over k_1 k_2}\right ) +{4 \over 7}\left({\bk_1\cdot\bk_2 \over k_1 k_2}\right )^2\,.
\een
\ees
Setting $a\rightarrow D_+(a)$ for the first order expressions used in Eq.~\eqref{eq:EdS_solution} is accurate to within $\mathcal{O}(2\%)$  \citep{Lee:2014gqa}. 
\para{EFT solutions} For the effective field theory case, one may similarly express the solutions in the form
\ben
&& \delta^{(n)}_c(\bk,a) \approx D_+^n(a)\int \frac{d^3\bk_1}{(2\pi)^3}\cdots \frac{d^3\bk_n}{(2\pi)^3}[\delta_D]_n F_n^{c}(\bk_1,\cdots,\bk_n;a)\delta^{(1)}(\bk_1)\cdots\delta^{(1)}(\bk_n)\,,\nn\\
&&  \bar{\theta}_c^{(n)}(\bk,a) \approx D_+^n(a)\int \frac{d^3\bk_1}{(2\pi)^3}\cdots \frac{d^3\bk_n}{(2\pi)^3}  [\delta_D]_n G_n^{c}(\bk_1,\cdots,\bk_n;a)\delta^{(1)}(\bk_1)\cdots\delta^{(1)}(\bk_n)\,,\nn\\
\een
noting that the kernels $F_n^c$ and $G_n^c$ have an implicit time dependence inherited from the time dependence of the source term, $\tau_\theta$. At first order the solution for $\delta_c$ is given by 
\ben
\delta^{(1)}_c(\bk,a)=-\xi(a) k^2 \delta^{(1)}(\bk,a)\,\quad {\text{where}}\quad \xi(a) = \int d\,a^{\prime}\, {\cal G}_{\delta}(a,a') d^2(a')\frac{D_+(a')}{D_+(a)}\,,
\een
while at second order the kernel $F_2^{c}$ is given by the sum of terms  $F_2^{\delta}+F_2^{e}+F_2^{\alpha\beta}$ as computed using the recipe given under Eq.~\eqref{eq:DeltaThetaExpansion} where
\bes\label{eq:F2_EFT}
\ben
&& F_2^{\delta}(\bk_1,\bk_2)=-\xi(a) E_\delta(\bk_1,\bk_2)\,, \\
&& F_2^{e}(\bk_1,\bk_2)=-\sum_{i=1}^3\epsilon_i(a) E_i(\bk_1,\bk_2) \,\,\text{with}\,\, \epsilon_i(a)= \int d\,a^{\prime}\, {\cal G}_{\delta}(a,a') e_i(a')\frac{D_+(a')^2}{D_+(a)^2},\\
&& F_2^{\alpha\beta}(\bk_1,\bk_2)=-\xi(a) E_{\alpha\beta}(\bk_1,\bk_2)\,.
\een
\ees
To give explicit expressions for the kernels $E_X$ requires knowledge of the time dependence of the coefficients of the stress-energy tensor $\tau_\theta$. As in \citep{bispec} we assume that $d^2, e_i\propto  D_+(a)^{m_d}$ for some specified value of $m_d$ (with constants of proportionality $\bar{d}^2,\bar{e}_i$, respectively). The results of \citep{bispec} showed robustness to different choices of $m_d$. Using the assumption of an EdS Universe one obtains the following expressions for the kernels,
\ben
&&E_\delta = M_d \bk_{12}^2 F_2(\bk_1,\bk_2) \quad \text{where} \,\, M_d \equiv {(m_d+1)(2m_d+7) \over (m_d+2)(2m_d+9)}\,,\nn\\
&&E_1(\bk_1,\bk_2) = \bk^2_{12}\,,\nn\\
&&E_2(\bk_1,\bk_2) = \bk^2_{12}\left [ {(\bk_1\cdot\bk_2)^2 \over k_1^2k^2} -{1\over 3}\right ]\,,\\
&&E_3(\bk_1,\bk_2)=-{1\over 6}\bk^2_{12} + \bk_1\cdot\bk_2 + {1\over 2}\left ({\bk_1\cdot\bk_2 \over k_1k_2} \right )^2 \left [ \bk_1^2 + \bk_2^2 \right ]\,,\nn\\
&& E_{\alpha\beta}(\bk_1,\bk_2) = {1\over 2m_d +9} \Big [ 2\beta(\bk_1,\bk_2)(\bk_1^2+\bk_2^2) \nn\\
&&\hspace{30mm}+ {2m_d+7 \over 2(m_d+2)}\left \{ \alpha(\bk_1,\bk_2)[\bk_2^2 +(m_d+2)\bk_1^2] +( {\bk_1 \leftrightarrow \bk_2}) \right \} \Big ]\,.\nn
\een
In addition with this choice of time dependence, the coefficients are given by $\xi(a)= \\2 \bar{d}^2 D_+^{m_d+1}/[(2m_d+7)(m_d+1)]$ and $\epsilon_i(a)=2\bar{e}_i D_+^{m_d+1}/[(2m_d+9)(m_d+2)]$. 

The expressions for $\bar{\theta}$ are similar, with $G_2^c=G_2^{\delta}+G_2^{e}+G_2^{\alpha\beta}$, with $G_2^{\delta}=(m_d+3)F_2^{\delta}$, $G_2^{e}=(m_d+3)F_2^{e}$, and 
\ben
G_2^{\alpha\beta}=-\xi(a)\frac{4(m_d+3)}{2 m_d+9}&& \Big [ 2\beta(\bk_1,\bk_2)(\bk_1^2+\bk_2^2) +\\
&& {3 \over 2(m_d+2)(m_d+3)}\left \{ \alpha(\bk_1,\bk_2)[\bk_2^2 +(m_d+2)\bk_1^2] +( {\bk_1 \leftrightarrow \bk_2}) \right \} \Big ]\,.\nn
\een
\para{Values of the counterterms} Using the assumption that $m_d=5/3$ (as suggested by self-similarity in an EdS Universe) the parameter $\xi$ was measured in \citep{bispec} to give a value of $(1.5 \pm 0.03) h^{-2}\text{Mpc}^2$ at redshift zero, while exact cancellation of divergences in the one loop SPT bispectra ensure,
\ben
&& { \epsilon_1 \over \xi}= {3466 \over 14091}\,, \quad { \epsilon_2\over \xi} = {7285 \over 32879}\,,
\quad {\epsilon_3 \over \xi}={41982 \over 52879}\,.
\label{eq:params}
\een
We shall assume these numerical values hold at other instances of $m_d$. Specifically, we will consider $m_d=1$ and $m_d=(1-n)/(3+n)$, where $n$ is the slope of the linear power spectrum, given approximately by $-3/2$ for large $k$.
\para{Bispectrum Predictions} The bispectrum for the density and velocity divergence perturbations is defined as
\ben
\langle \delta(\bk_1)\delta(\bk_2)\delta(\bk_3)\rangle\equiv
(2\pi)^3 \delta_{\rm 3D}(\bk_1+\bk_2+\bk_3)B_{\delta}(k_1,k_2,k_3)\,,\\
\langle \bar{\theta}(\bk_1)\bar{\theta}(\bk_2)\bar{\theta}(\bk_3)\rangle\equiv
(2\pi)^3 \delta_{\rm 3D}(\bk_1+\bk_2+\bk_3)B_{\bar{\theta}}(k_1,k_2,k_3)\,.
\een
The tree-level SPT bispectra are computed using two first order and one second order perturbations, $\langle \delta^{(1)}\delta^{(1)}\delta^{(2)} + \text{cyclic}\rangle$, to give
\ben\label{eq:SPT_tree}
&&B_{\delta}^{\text{SPT}}(k_1,k_2,k_3)=2 F_2(\bk_1,\bk_2) P(k_1)P(k_2) + \text{cyclic}\,, \label{eq:SPT_Bisp}\\
&&B_{\bar{\theta}}^{\text{SPT}}(k_1,k_2,k_3)=2 G_2(\bk_1,\bk_2) P(k_1)P(k_2) + \text{cyclic}\,,\label{eq:SPT_Bisp2}
\een
where $P(k)$ denotes the tree level power spectrum.

The leading EFT contribution may be evaluated using $\langle \delta^{(1)}_c\delta^{(1)}\delta^{(2)}+5\text{ permutations}\rangle$ $+\langle  \delta^{(1)}\delta^{(1)}\delta^{(2)}_c + 2\text{ cyclic}\rangle $, and similarly for the velocity divergence. The resulting density bispectra are given by (where we suppress explicit dependence on wavenumbers for simplicity)
\ben\label{eq:B_EFT_contribs}
&& B_\delta^{\text{EFT}}=B_{\delta_c^{(1)}}+B_{\delta_c^{(2)},\delta}+B_{\delta_c^{(2)},e}+B_{\delta_c^{(2)},\alpha\beta}\,;\\
&& B_{\delta^{(1)}}=-\xi(a) \Big[2 F_2(\bk_1,\bk_2)(k_1^2+k_2^2) P(k_1)P(k_2) + 5\text{ permutations}\Big]\,,
\een
while the other expressions are given by Eq.~\eqref{eq:SPT_Bisp} with the replacement of the kernel $F_2$ by the appropriate formula in Eq.~\eqref{eq:F2_EFT} indicated by the subscript. The first term was absent from the expressions presented in \citep{bispec}.

Similarly, for the velocity divergence the EFT bispectra may be written in the form,
\ben\label{eq:veldiv_EFT}
&& B_{\bar{\theta}}^{\text{EFT}}=B_{\bar{\theta}_c^{(1)}}+B_{\bar{\theta}_c^{(2)},\delta}+B_{\bar{\theta}_c^{(2)},e}+B_{\bar{\theta}_c^{(2)},\alpha\beta}\,;\\
&& B_{\bar{\theta}^{(1)}}=-(m_d+2)\xi(a) \Big[2 G_2(\bk_1,\bk_2)(k_1^2+k_2^2) P(k_1)P(k_2) + 5\text{ permutations}\Big]\,,
\een
with the remaining terms obtained by replacing $G_2$ in Eq.~\eqref{eq:SPT_Bisp2} with the indicated kernel for $\bar{\theta}_c^{(2)}$. 
\para{Squeezed limit comparisons using tree level SPT} While comparison of the SPT bispectrum to the leading order EFT contribution should be done using the 1-loop SPT formula, we shall in this paper perform our computations using the tree level SPT results of Eqs.~\eqref{eq:SPT_Bisp} and~\eqref{eq:SPT_Bisp2}. The justification for this simplification is that our focus in this paper is on the suqeezed bispectrum, for which the tree level prediction agrees to $\mathcal{O}(5\%)$ with the one-loop result out to $k\approx 0.1 h/\text{Mpc}$ and to within $\mathcal{O}(10\%)$ in the range $k\in[0.1,0.2]h/\text{Mpc}$ \citep{Thesis} at redshift zero. In \citep{Thesis} it was also conjectured that discrepancies between the SPT predictions and measurements from simulations is most likely due to non-perfect fluid terms which are the primary focus of this paper.
%
%
\section{Primordial non-Gaussianity in EFT}\label{sec:PNG}
The Planck surveyor has put very tight constraint on primordial non-Gaussianity \citep{Planck}.
However, future CMB missions may not be able to significantly lower the bound on primordial non-Gaussianity.
On the other hand, galaxy redshift surveys are currently entering a new era in terms of survey volume, number of galaxies observed, 
as well as the range of redshift probed. This will allow very high signal-to-noise for the detection of higher-order
statistics \citep{shapes}, and also lead to further tightening of the constraints on primordial non-Gaussianity or, indeed, possible detection.

Our discussion of the SPT and EFT solutions for the density and velocity divergence perturbations in Sec.~\ref{sec:eftoflss} was implicitly predicated on the assumption of Gaussian initial conditions. In the non-Gaussian case, a non-zero primordial bispectrum of the gravitational potential, $B_\phi$ allows for a further degree of freedom with which to construct the effective stress-energy tensor. More particularly, the potential, $\phi$, is expanded about a Gaussian field $\phi_g$, such that for $k\neq 0$ \citep{primordial}
\ben\label{eq:PhiExpansion}
\phi(\bk)=\phi_g(\bk)+f_{\text{NL}}\int \frac{d^3 p}{(2\pi)^3} K_{\text{NL}}({\bf p},\bk-{\bf p})\left[ \phi_g({\bf p})\phi_g(\bk-{\bf p})\right]\,,
\een 
where $f_{\text{NL}}$ labels the amplitude of the bispectrum of the form,
\ben
B_{\phi}(k_1,k_2,k_3)=2 f_{\text{NL}} K_{\text{NL}}(\bk_1,\bk_2) P_{\phi}(k_1)P_{\phi}(k_2) + \text{cyclic}.
\een
Noting the relation between the gravitational potential and the density fluctuation, $\delta = M(k,a) \phi$, where $M(k,a)=2 k^2 /(3 {\cal H}^2 \Omega_M)$, as in Eq.~\eqref{eq:Poisson}, the contribution to the total matter bispectrum is given by
\ben
B_\delta(k_1,k_2,k_3)\supset B_{\text{PNG}}^{\text{SPT}}(k_1,k_2,k_3)\equiv M(k_1,a)M(k_2,a)M(k_3,a)B_\phi(k_1,k_2,k_3)\,.
\een
where we suppress the redshift dependence in our notation for the bispectrum. 

In the squeezed limit, scalar contributions from $K_{\text{NL}}$ may be expressed in the form
\ben
K_{\text{NL}}(\bk_1,\bk_2)\xrightarrow{k_3\ll k_1,k_2} \frac{B_{\phi}(k_1,k_2,k_3)}{4 f_{\text{NL}}P_{\phi}(k_1)P_{\phi}(k_2)}\approx a_0 \left(\frac{k_3}{k}\right)^\Delta\,,
\een
for some constant $a_0$ and scale factor $\Delta$. This in turn dictates that the long mode contribution to Eq.~\eqref{eq:PhiExpansion} can be written $f_{\text{NL}}\psi(\bk)$, where $\psi(\bk)=(k/\mu)^{\Delta}\phi_g(\bk)$ for some arbitrary scale\footnote{For simplicity, in this paper we choose $\mu=1 \,h/\text{Mpc}$.} $\mu$. This contribution modulates the short wavelength contribution. 

For the purposes of this paper, we consider the local, equilateral and quasi-single-field models of inflation for which the scaling dimension $\Delta$ is given by $\{0,2,1\}$, respectively, and for which the bispectra are given by
\bes\label{eq:PNG_models}
\ben
\label{eq:local}
&& B^{\rm loc}_{\phi}(\bk_1,\bk_2,\bk_3)= 2f^{\rm loc}_{\rm NL}\left(P_{\phi}(k_1)P_{\phi}(k_2)+\text{cyclic} \right)\,, \\
&& B^{\rm eq}_{\phi}(\bk_1,\bk_2,\bk_3) = 162 f^{\rm eq}_{\rm NL} A^2_{\phi} {1\over k_1k_2k_3 K^3}\,, \label{eq:equil}\\
&& B^{\rm qsf}_{\phi}(\bk_1,\bk_2,\bk_3) = 18\sqrt{3} f^{\rm qsf}_{\rm NL} A^2_{\phi} {1\over k_1k_2k_3 K^3} 
{N_{\nu}(8\kappa)\over \sqrt{\kappa} N_{\nu}(8/27)}\,,
\label{eq:qsf}
\een
\ees
where, we have defined $K={k_1+k_2+k_3}$ and $\kappa = k_1k_2k_3/K^3$, and expressed the power spectrum of the gravitational potential in the form $P_\phi(k)=A_\phi/k^3 (k/k_s)^{n_s-1}$, where $k_s$ is a pivot scale and $n_s$ represents the scalar tilt. $N_{\nu}$ denotes the Neumann function of order $\nu$,
which is taken to be $\nu={1/2}$ for definiteness, though the index $\nu$ can take a wider range of allowed values. 
\para{EFT contributions} The long mode modulation of the gravitational potential gives the extra contributions to the effective stress energy tensor given in Eq.~\eqref{eq:ansatz_noPNG} \citep{primordial},
\ben\label{eq:ansatz_PNG}
\tau_\theta\supset - \fNL\left(g(\triangle \Psi - \partial_i(\delta \partial^i \Psi))+g_1 \triangle(\Psi \delta)+g_2\partial_i \partial_j(\Psi s^{ij})\right)\,,
\een
for coefficients $g,g_1,g_2$ and where $\Psi(\bx)$ is given by $\psi(\bq(\bx))$ at the Lagrangian position, $\bq(\bx)$. This results in the following contributions to the density bispectrum,
\ben\label{eq:PNG_SPT_EFT}
&& B_{\text{PNG}}^{\text{EFT}}=B_{\text{PNG}}^{\text{SPT}}-\fNL\left( B_{\text{PNG}}^{g(1)}+B_{\text{PNG}}^{g(2)}+B_{\text{PNG}}^{g_1}+B_{\text{PNG}}^{g_2}+B_{\text{PNG}}^{\alpha\beta}\right)\,;\\
&& B_{\text{PNG}}^{g(1)}= \gamma(a) (k_1^2+k_2^2+k_3^2) B_{\text{PNG}}^{\text{SPT}}\,, \nn\\
&& B_{\text{PNG}}^{g(2)}=\gamma(a) \left[\frac{m_g (2 m_g+5)}{(m_g+1)(2 m_g+7)}\right]\left(\bk_{12}^2 \frac{\bk_1.\bk_2}{k_2^2}-\bk_{12}.\bk_1\right) P_{1\psi}(k_1) P(k_2)+5 \text{ permutations}      \,, \nn\\
&& B_{\text{PNG}}^{g_1}=\gamma_1(a) \bk_{12}^2 P_{1\psi}(k_1) P(k_2)+5 \text{ permutations} \,,\nn\\
&& B_{\text{PNG}}^{g_2}=\gamma_2(a) \Big[\frac{(\bk_{12}.\bk_2)^2}{k_2^2}-\frac{\bk_{12}^2}{3} \Big] P_{1\psi}(k_1) P(k_2)+5 \text{ permutations}\,,\nn\\
&& B_{\text{PNG}}^{\alpha\beta}= 4 \gamma(a) \Big[\frac{4}{2 m_g+7}\beta(\bk_1,\bk_2)+\nn\\
&&\hspace{10mm}\frac{2 m_g+5}{(m_g+1)(2m_g+7)}((m_g+1)\alpha(\bk_1,\bk_2)+\alpha(\bk_2,\bk_1) \Big] k_1^2 P_{1\psi}(k_1) P(k_2)+5 \text{ permutations}\,,\nn
\een
where $P_{1\psi}(k)\equiv (k/\mu)^\Delta P(k)/M(k)$.
Here we have assumed that $g(a)\propto D_+(a)^{m_g}$, and have solved in the limit of an exact EdS Universe -- extrapolating the result by the replacement $a\rightarrow D_+(a)$. In addition the time dependent factors are given by $\gamma(a)=-\int da^{\prime}{\cal G}_{\delta}(a,a^{\prime}) \\ g(a^{\prime})D_{+}(a^{\prime})/D_{+}(a)$, and similarly for $\gamma_i(a)$ (with $g$ replaced by $g_i$ in the integrand).
%
%
\section{Redshift Space Distortions}\label{sec:RSD}
Next we consider the case of galaxy bispectrum in redshift space~\citep{MVH1, Scocci1, Verde1, Scocci2, Scocci3, HMV2}. An extra complication arises when the redshift space effects are accounted for \citep{Jack72}, with the galaxy survey measuring the redshift corresponding to \textit{both} the Hubble flow and the peculiar velocity, ${\bf v}$ along the line of sight, ${\hat \bx}_{\parallel}$. The peculiar velocities result in a displaced radial position
\ben
{\bf s}= {\bx} + \frac{{\bf v}.{\hat \bx}_{\parallel}}{a H}{\hat \bx}_{\parallel}\,,
\een
which, in turn, distorts the measured density field. The relation between the redshift space matter overdensity, $\delta_s$, and the real space $\delta$, was inferred by Scoccimarro \citep{Scoccimarro:2004tg} by noting that the mapping must conserve mass, such that
\ben
\delta_s(\bk)=\delta(\bk)+\int d^3 x e^{- i \bk.{ \bx}_{}}\left[ \exp\left(-\frac{i (\bk.{\hat \bx}_{\parallel})({\bf v}.  {\hat \bx}_{\parallel}      )}{H}\right)\right][1+\delta(\bk)]\,.
\een
To generalise this expression to galaxies from matter, one must also account for the biasing of tracers.
There is a rich literature of estimation from galaxy survey~\citep{FFFS,SFFF,JingBorner,GazScoc,WangYang,Marin,KR}. While we will
primarily be interested in the squeezed limit and correction from EFT counter-terms, we first shall recapitulate the standard perturbation theory expressions presented in those works. The expressions presented here will be valid in the distant observer approximation. However, we note that future surveys will probe a 
considerable fraction of the sky. A 3D approach has been developed which 
uses spherical-Bessel transform, and has been used recently to compute the redshift power spectrum 
\citep{PM13,galaxy_red, Bernardeau_wideangle}.

The effect of redshift-space distortions (RSD) is to mix peculiar velocity statistics with the statistics of the density contrast. Interpretation and analysis of results from spectroscopic surveys therefore
depends on our ability to model such complications.
We will denote the redshift-space density contrast of halos as $\delta_{h,s}$, which -- similar to their real-space counterparts -- can be expressed in terms of redshift-space kernels $Z_n(\bk_1,\cdots,\bk_n)$ and the bias parameters $b_k$ (c.f. Eq.~\eqref{eq:EdS_solution}):
\ben
\delta_{h,s}(\bk,\tau) = \sum^{\infty}_{n=1} D^n_+(\tau)  \int \frac{d^3 \bk_1}{(2\pi)^3}\cdots \int \frac{d^3\bk_n}{(2\pi)^3}\,[\delta_D]_n Z_n(\bk_1,\cdots,\bk_n)\;
\delta^{(1)}(\bk_1)\cdots\delta^{(1)}(\bk_n).
\een
The lower order kernels $Z_1(\bk_1)$ and $Z_2(\bk_1,\bk_2)$ in redshift-space are given by the following expressions \citep{red_roman}:
\bes
\ben
\label{eq:Z1}
&& Z_1(\bk_i) \equiv (b_1+f\mu_i^2) \\
&& Z_2(\bk_1,\bk_2)\equiv b_1\left [F_2(\bk_1,\bk_2)+ 
{1 \over 2}f\mu k\left ({\mu_1 \over k_1} + {\mu_2 \over k_2} \right )\right ]
+ f\mu^2 G_2(\bk_1,\bk_2\nn) \label{eq:Z2}\\
&& \quad\quad\quad + {1 \over 2}f\mu k \mu_1\mu_2\left ({\mu_1 \over k_1} + {\mu_2 \over k_2} \right )
+{b_2 \over 2} +{b_{s^2} \over 2}S_2(\bk_1,\bk_2)\,,
\een
\ees
where, here and throughout, we use the expression for the direction cosines:
\ben\label{eq:RSD_angles}
&& \mu =  {\hat \bx}_{\parallel} \cdot {\hat \bk};\quad\quad \mu_i ={\hat \bx}_{\parallel}\cdot{\hat\bk}_i\,,
\een
with the comoving separation separated into components that are parallel and perpendicular to the 
line of sight $\bx=\bx_{\parallel}+\bx_{\perp}$.
The direction cosines are related by
\ben
&& \bk=\bk_1+\bk_2; \quad \mu k \equiv (\mu_1 k_1 + \mu_2 k_2). 
\een
The kernel  $Z_2(\bk_1,\bk_2)$ defined in Eq.(\ref{eq:Z2}) depends on both $F_2(\bk_1,\bk_2)$ and $G_2(\bk_1,\bk_2)$ which implies that the squeezed limit of redshift
space bispectrum will depend on the squeezed limits of $\delta$ and $\theta$ bisepctrum \citep{Munshi2}.
To relate the halo density contrast $\delta_h$ and the underlying contrast $\delta$
we use a deterministic bias: $\delta_h= \sum_k b_k {\delta^k / k!}$, noting that the results can be simply extended to  include  scale-dependent bias, and other complications. 

For these computations we have utilised a more general form of the 
{\em symmetrized} second order kernels $F_2(\bk_1,\bk_2)$ and $G_2(\bk_1,\bk_2)$ \citep{Munshi2}:
\ben
&& F_2(\bk_1,\bk_2) = {1\over 2}{(1+\epsilon)} + {1 \over 2}{\mu_{12}}\left ( {k_1\over k_2}+{k_2\over k_1}\right )
+{1\over 2}{(1-\epsilon)}\mu^2_{12}; \quad \mu_{12}={\hat \bk_1\cdot \hat \bk_2}.
\label{eq:F2}
\een
where the parameter, $\epsilon$, takes the value $3/7$ for an Einstein de-Sitter Universe. The kernel $G_2(\bk_1,\bk_2)$ has similar functional form; to avoid confusion we will use the parameter $\epsilon^{\prime}$ 
 with $\epsilon^{\prime}=-1/7$ relevant for the SPT kernel. This parametrisation is useful as similar calculations may be performed using e.g.
first order Lagrangian Perturbation Theory (LPT), also known as the Zel'dovich Approximation (ZA;
\citep{MunshiStarobinsky,MunshiSahniStarobinsky}),  with the parameter in this case taking the value $\epsilon={0}$.

\para{RSD power spectrum and bispectrum}
In redshift-space the halo power spectrum takes the form \citep{review}
\ben
&&  \langle\delta_{h,s}(\bk_1)\delta_{h,s}(\bk_2)\rangle \equiv (2\pi)^3\delta_{\rm 3D}(\bk_1+\bk_2) P_{h,s}(k_1) ; \\
&& P_{h,s}(k) =  b_1^2(1+ b_1^{-1}f\mu_k^2)^2\,P_{}(k).
\een
Similarly the redshift-space halo bispectrum, $B_{h,s}$, is defined via \cite{red_roman},
\bes
\ben
\label{eq:bispec_red}
&& \langle\delta_{h,s}(\bk_1)\delta_{h,s}(\bk_2)\delta_{h,s}(\bk_3)\rangle_c \equiv
(2\pi)^3 \delta_{\rm 3D}(\bk_1+\bk_2+\bk_3)B_{h,s}(\bk_1,\bk_2,\bk_3);\\
\label{eq:fog1}
&& B_{h,s}(\bk_1,\bk_2,\bk_3) = D_{\rm FoG}
[2P(\bk_1){\rm Z}_1(\bk_1)P(\bk_2){\rm Z}_1(\bk_2){\rm Z}_2 (\bk_1,\bk_2) + {\text{cyclic}}].\\
&& D_{\rm FoG}(k_1,k_2,k_3,\sigma_{\rm FoG}[z])
=(1+[k_1^2 \mu^2_{1} + k_2^2 \mu^2_{2} + k_3^2 \mu^2_{3}]^2 \sigma^2_{\rm FoG}[z]/2)^{2}.
\label{eq:fog2}
\een
\ees
At small scales the {\em Fingers-of-God} (FoG) effect can dominate clustering in redshift space.
The FoG effect arises as a result of random peculiar velocities of galaxies within virialised collapsed objects.
The effect of peculiar velocity is an incoherent contribution and results in a suppression
of the clustering amplitude at high $k$ \citep{Jack72}.  This is distinct from the requirement to resum large-scale random motions which are not fully accounted for within the standard perturbation theory framework. The effect of these bulk flows (once resummed) is to dampen the effect of acoustic oscillations. Within the EFT framework, the FoG effect is accounted for by the counterterms, while the large-scale bulk flow can be modelled by re-writing the SPT expressions in the Lagrangian perturbation theory \cite{resum,Matsubara,Vlah,Vlah:2015zda,delaBella:2017qjy}. Nevertheless, following the standard nomenclature in the literature we denote the multiplicative factor, which accounts for the large-scale un-resummed bulk velocities, as the Fingers-of-God contribution in Eq.(\ref{eq:fog1})-Eq.(\ref{eq:fog2}).

\para{SPT contributions}
Following \citep{Thesis} we will group the various configurations to the bispectrum in Eq.(\ref{eq:bispec_red}) as follows:
\bes
\ben
&& {B} = {B}_{\rm SQ_1}+{B}_{\rm SQ_2}+{B}_{\rm NLB}+{B}_{\rm FoG}; \\
&& {B_{SQ_1}} = b_1^3\sum \beta^{ i-1}B_{{SQ_1},i}; \quad {B_{SQ_2}} = b_1^3\beta \sum \beta^{ i-1}B_{{\rm SQ_2},i}; \quad \beta = {f/b_1}; \\
&& {B_{\rm NLB}} = b_1^2b_2\beta \sum \beta^{ i-1}B_{{\rm SQ_2},i}. \\
&& {B}_{\rm FoG} = b_1^4 \beta[ B_{\rm FoG_1}+\beta (B_{\rm FoG_2} +B_{\rm FoG_3}) +
\beta^2(B_{\rm FoG_4}+B_{\rm FoG_5}) + \beta^3 B_{\rm FoG_6}].
\een
\ees
The first contributions ${B}_{\rm SQ_1}$ listed above, represents linear squashing and depends on the kernel $F_2(\bk_1,\bk_2)$.
The linear Kaiser effect \citep{Kaiser87} represents the coherent distortion due to  the peculiar velocity
along the line of sight. The linear growth rate controls its magnitude. At the level of power spectrum it leads to
an enhancement of the power spectrum amplitude at small $k$.
The second order squashing terms  ${B}_{\rm SQ_2}$ depends on the kernel $G_2(\bk_1,\bk_2)$.
Nonlinear biasing is represented by the  $B_{\rm NLB}$ terms and depend on the second order biasing coefficient $b_2$. Finally $B_{\rm FOG}$ represents the additive FoG effect which needs to be included in addition
to the multiplicative factor $D_{\rm FOG}$ introduced above. 
The real-space expressions can be recovered by taking the limit $f\rightarrow 0$ in Eq.(\ref{eq:Z1})-Eq.(\ref{eq:Z2}).

More explicitly -- from \citep{Thesis} -- the ``linear squashing" terms (SQ$_1$)  are given by 
\begin{subequations}
\ben
\label{eq:LOS1}
&& B_{\rm SQ1_1}(\bk_1,\bk_2,\bk_3) = 2[F_2(\bk_1,\bk_2)P_{}(k_1)P_{}(k_2)+{\text{cyclic}}]\,,\\
&& B_{\rm SQ1_2}(\bk_1,\bk_2,\bk_3) = 2[(\mu_1^2+\mu_2^2)F_2(\bk_1,\bk_2)P_{}(k_1)P_{}(k_2)+{\text{cyclic}}]\,,\\
&& B_{\rm SQ1_3}(\bk_1,\bk_2,\bk_3) = 2[\mu_1^2\mu_2^2F_2(\bk_1,\bk_2)P_{}(k_1)P_{}(k_2)+{\text{cyclic}}]\,,
\label{eq:LOS3}
\een
\end{subequations}
and the ``second order squashing" terms (SQ$_2$) are written 
\begin{subequations}
\ben\label{eq:LOS2}
&& B_{\rm SQ2_1} = 2[\mu^2G_2(\bk_1,\bk_2)P_{}(k_1)P_{}(k_2)+{\text{cyclic}}]\,,\\
&& B_{\rm SQ2_2} = 2[\mu^2(\mu_1^2+\mu_2^2)G_2(\bk_1,\bk_2)P_{}(k_1)P_{}(k_2)+{\text{cyclic}}]\,,\\
&& B_{\rm SQ2_3} = 2[\mu^2\mu_1^2\mu_2^2G_2(\bk_1,\bk_2)P_{}(k_1)P_{}(k_2)+{\text{cyclic}}]\,.
\een
\end{subequations}

The contributions corresponding to the ``non-linear bias'' (NLB) terms and ``Fingers-of God" (FoG) terms are independent of kernels $F_2(\bk_1,\bk_2)$ and
$G_2(\bk_1,\bk_2)$, so do not depend on $\epsilon$ or $\epsilon^{\prime}$. These contributions do not receive corrections (at leading order) from EFT. We refer the reader to \citep{Thesis} for full expressions of $B_{\text{NLB}_i}$ ($i\in [1,3]$) and $B_{\text{FOG}_i}$ ($i\in[1,6]$).

\para{EFT contributions} The corrections to the SPT redshift space distortion expressions due to the UV physics encapsulated within the EFT framework at leading order may be inferred from Sec.~\ref{sec:eftoflss}, via the replacement of $F_2$ and $G_2$ by $F_2^c$ and $G_2^c$, respectively within the expressions for $\text{SQ1}$ and $\text{SQ2}$.

The bispectrum predicted by modified gravity theories in general can be very different from the standard 
$\Lambda$CDM predictions which we have considered here \citep{BorisovJain,TT,Hector,Bartolo,Bellini}.
Using simulations of MG theories, the EFT parameters can be estimated
which can be used to obtain results similar to what is presented in this paper.  

\section{Squeezed Limits, Consistency Relations \& Position Dependent Power Spectra}\label{sec:consistencyRels}
\subsection{Consistency Relations and the Position Dependent Power Spectrum}
Consistency relations connect the correlations between different orders in the squeezed limit \citep{KNPR}. Equal-time relations vanish in the soft limit for a uniform gravitational field as (1) the equivalence principle ensures that the effect of the zero mode and first spatial gradient of the long wavelength mode can be locally eliminated, and (2) the short-scale modes are uniformly displaced. As emphasised in \citep{equal,Baldauf:2011bh} the relations will not vanish for a non-uniform gravitational field, or in the presence of a non-Gaussian initial field. \citep{equal} derived the consistency relation for the $n+1$-point function which is valid in the mildly non-linear regime,
\ben\label{eq:consistencyrelation}
\langle \delta(\bq)\delta(\bk_1)\cdots \delta(\bk_n)\rangle^{\text{av}}_{q\rightarrow 0}=P(q) \Big[1-\sum_{i=1}^n {1\over 3}{\partial \over \partial \ln k_i} +{13 \over 21}{\partial \over \partial \ln D_+(a)} \Big] \langle \delta(\bk_1)\cdots \delta(\bk_n)\rangle\,,
\een
where the superscript $``\text{av}"$ indicates an averaging over the angle between the long and short wavelength modes.
Going beyond the mildly non-linear regime necessitates either a phenomenological model, such as the halo model \cite{halo}, to probe non-perturbative scales, a fitting formula approach (e.g. \citep{GilMarin:2011ik}), or the use of frameworks such as the effective field theory of large scale structure investigated in this paper. The last term in the consistency relation is only valid in the quasilinear regime which motivates going beyond the quasilinear regime -
which also motivates the use of EFT to derive consistency relations that are valid beyond the perturbative regime.
The consistency relations or equivalently the response functions have also been checked using simulations \citep{BS,WSCK}
The form of these expressions in redshift space was investigated in \citep{const1} who derived the extension of the relation to account for the velocity component along the radial direction, where they were studied at leading order SPT, and using the non-perturbative approximation to investigate the non-linear regime (but only in the one-dimensional case). As a probe of non-uniformity of the gravitational field, consistency relations have been suggested as a testbed for theories of modified gravity \citep{Creminelli1}.
The presence of primordial non-Gaussianity also invalidates the standard formulae for the consistency relations due to the introduction of a correlation between the long and short wavelength modes.
\para{Position Dependent Power Spectrum and Integrated Bispectrum} The integrated bispectrum is defined by first dividing a survey into $N_s$ subvolumes centred at $\br_L$; within each subvolume one computes the power spectrum and average overdensity, denoted $P(\bk,\br_L)$ and $\bar{\delta}_L$, respectively. The appropriate three point function is given by the expectation value of the product of these quantities. However, due to non-isotropy of the window functions defining the subvolumes, the position dependent power spectrum, $P(\bk,\br_L)$, may depend on the orientation $\hat{\bk}$. Thus one averages over orientations to give the expression for the integrated bispectrum,
\ben\label{eq:iBdef}
{iB}(k)\equiv \int \frac{d^2 \hat{k}}{4\pi}\langle P(\bk,\br_L)\bar{\delta}(\br_L)\rangle_{N_s}\,,
\een
where the subscript $N_s$ is used to emphasise that the expectation is taken over all subvolumes. The integrated bispectrum is related to the linear response function, $d\ln P(k)/d\bar{\delta}$, as can be demonstrated by taking the Taylor expansion of $P(\bk,\br_L)$ in powers of $\bar{\delta}(\br_L)$, to establish that at leading order \citep{komatsu}
\ben
{iB}(k)\approx {d\ln P(k) \over d\bar{\delta}}\Big|_{\bar{\delta}=0} P(k) \sigma_L^2 \quad \text{where}\quad \sigma_L^2\equiv \langle \bar{\delta}(\br_L)^2\rangle_{N_s}\,.
\een
The explicit relationship between the integrated bispectrum and the bispectrum is given by \citep{komatsu}
\ben
{iB}(k) &=& \frac{1}{V_s^2}\int \frac{d^2 \hat{k}}{4\pi}\int \frac{d^3 q_a}{(2\pi)^3}\int \frac{d^3 q_b}{(2\pi)^3} \Big[ B(\bk-\bq_a,-\bk+\bq_a+\bq_b,-\bq_b)\nn\\
 && \hspace{60mm} \times W_L(\bq_a)W_L(\bq_b)W_L(-\bq_a-\bq_b)\Big]\,,
\een
where the window functions, $W_L$, largely constrain the $q_i$ integrals to within the subvolumes (of volume $V_s$). For $k$ larger than the subvolume scale $\sim 1/V_s^{1/3}$ the integrals are, therefore, dominated by contributions from squeezed bispectrum configurations, with $q_a,q_b\ll k$. Thus, in this paper we will ignore the exact form of the window functions and focus instead on the (computationally simpler) angle-averaged squeezed bispectrum given by (noting that $k_3\equiv q_b$)
\ben
B^{\text{sq}}(k,k_3)  \stackrel{q_a, q_b\ll k}{\equiv} \int \frac{d^2 \hat{k}}{4\pi} \int \frac{d^2 \hat{q}_a}{4\pi}\int \frac{d^2 \hat{q}_b}{4\pi}B(\bk-\bq_a,-\bk+\bq_a+\bq_b,-\bq_b)\,.
\een
Thus, by comparison to Eq.~\eqref{eq:consistencyrelation} it is apparent that the angle-averaged squeezed bispectrum studied in this paper is equivalent to probing the consistency relations and can be used as a proxy for the integrated bispectrum.

\subsection{Squeezed limit of the EFT density bispectrum}\label{sec:Density}
\subsubsection{General ($3$D) case}
In Sec.~\ref{sec:eftoflss} we described the effective field theory of large scale structure and its particular application for the computation of the density and velocity divergence bispectra. Here we largely follow the description in  \citep{komatsu,Thesis} to compute the squeezed limit of the expressions described for the density.

\begin{figure}
\centering
\text{\hspace{1cm}Gravity Induced Non-Gaussianityin 3D - $\delta$}\par\smallskip
\includegraphics[width=0.45\textwidth]{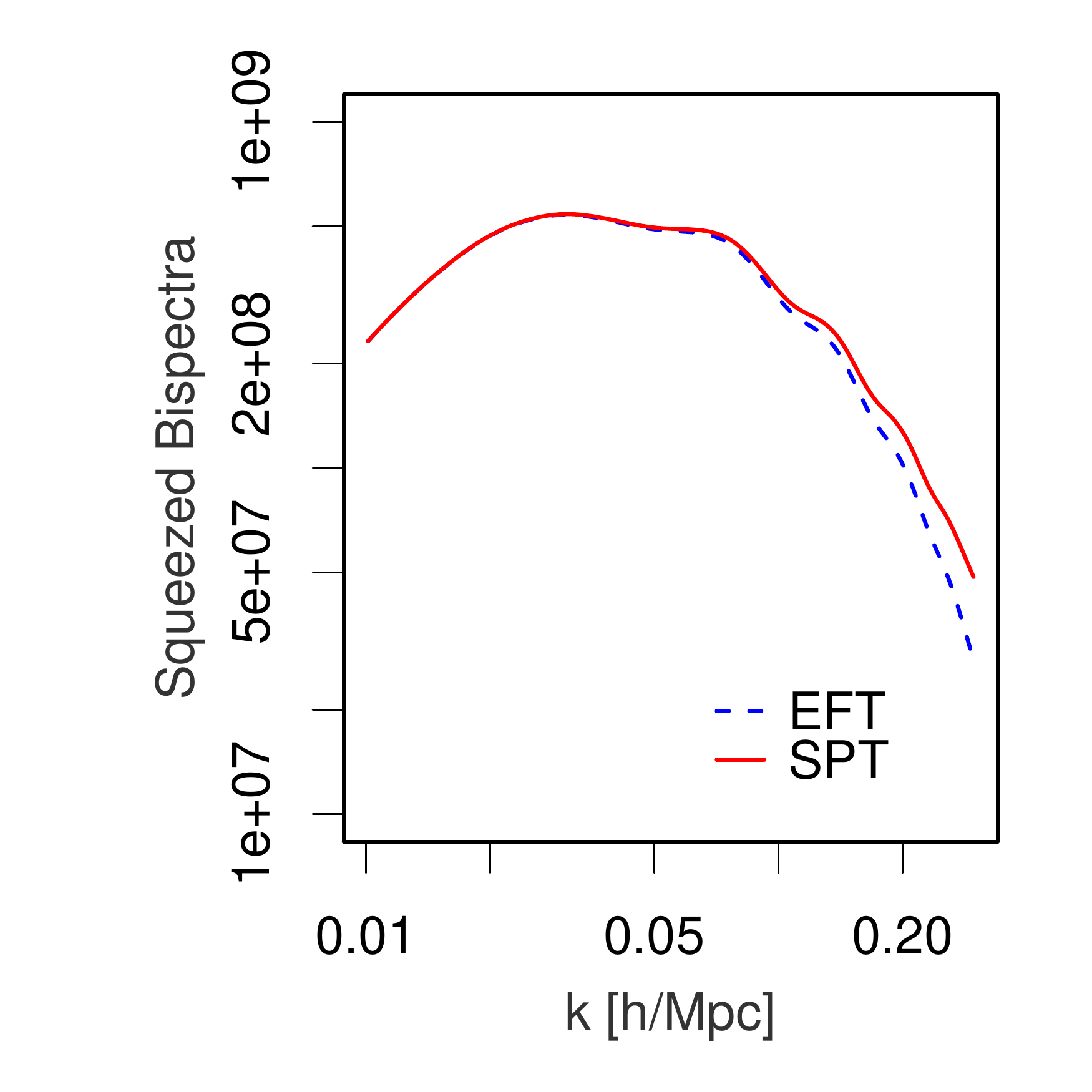}
\includegraphics[width=0.45\textwidth]{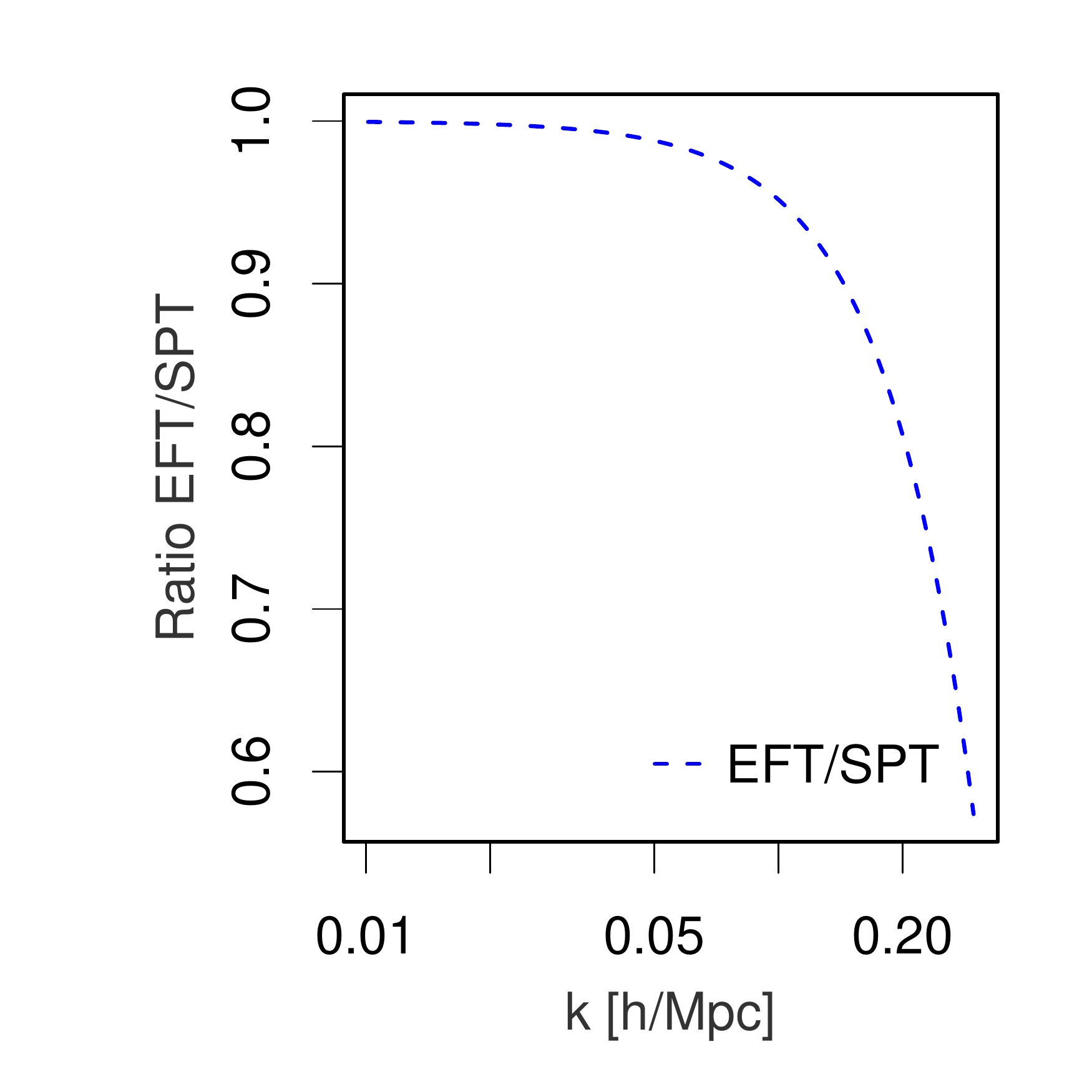}
\caption{(Left panel) Squeezed matter bispectrum $B_\delta(\bk,-\bk,-\bq_b)$ for $q_b\equiv k_3=k/10$, for the both SPT (red, solid) and EFT (blue, dashed). (Right panel) Ratio of the EFT and SPT squeezed bispectra. 
Corrections to the usual consistency relation become important for $k\gtrsim 0.05\,\rm {h/Mpc}$ [see Eq.(\ref{eq:sq_EDS})-Eq.(\ref{eq:sq_EFTDelta}) for analytical expressions.]}
\label{fig:bSPT_bEFT}
\vspace*{2em}
\end{figure}
\begin{figure}
\centering
\text{\hspace{1cm}Gravity Induced Non-Gaussianity in 3D - $\delta$}\par\smallskip
\includegraphics[width=0.5\textwidth]{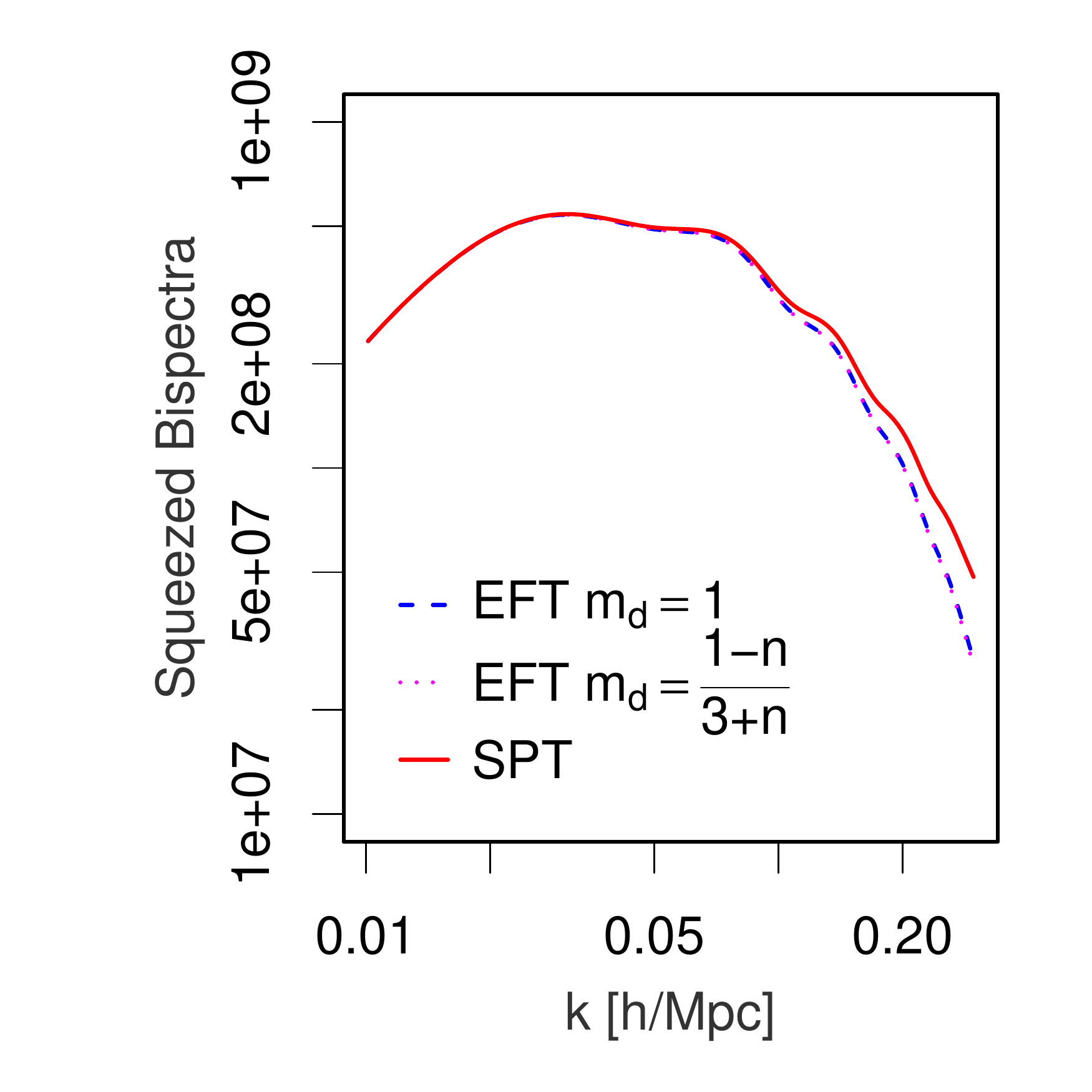}
\caption{Two different approximations for $m_d$ are compared for the scaling dependence of the counterterms.
The different values of $m_d$ correspond to those typically considered in the literature, namely  $m_d={(1-n)/(n+3)}$ and $m_d=1$.
The first value corresponds to EdS Universe and for $n=-3/2$ can approximate $\Lambda$CDM, while most of our numerical results have been evaluated using $m_d=1$ which corresponds to the time-dependence of the 1-loop terms.  }
\label{fig:bSPT_approx}
\vspace*{2em}
\end{figure} 
\begin{figure}
\centering
\text{\hspace{1cm}Gravity Induced Non-Gaussianity in 3D - $\delta$}\par\smallskip
\includegraphics[width=0.45\textwidth]{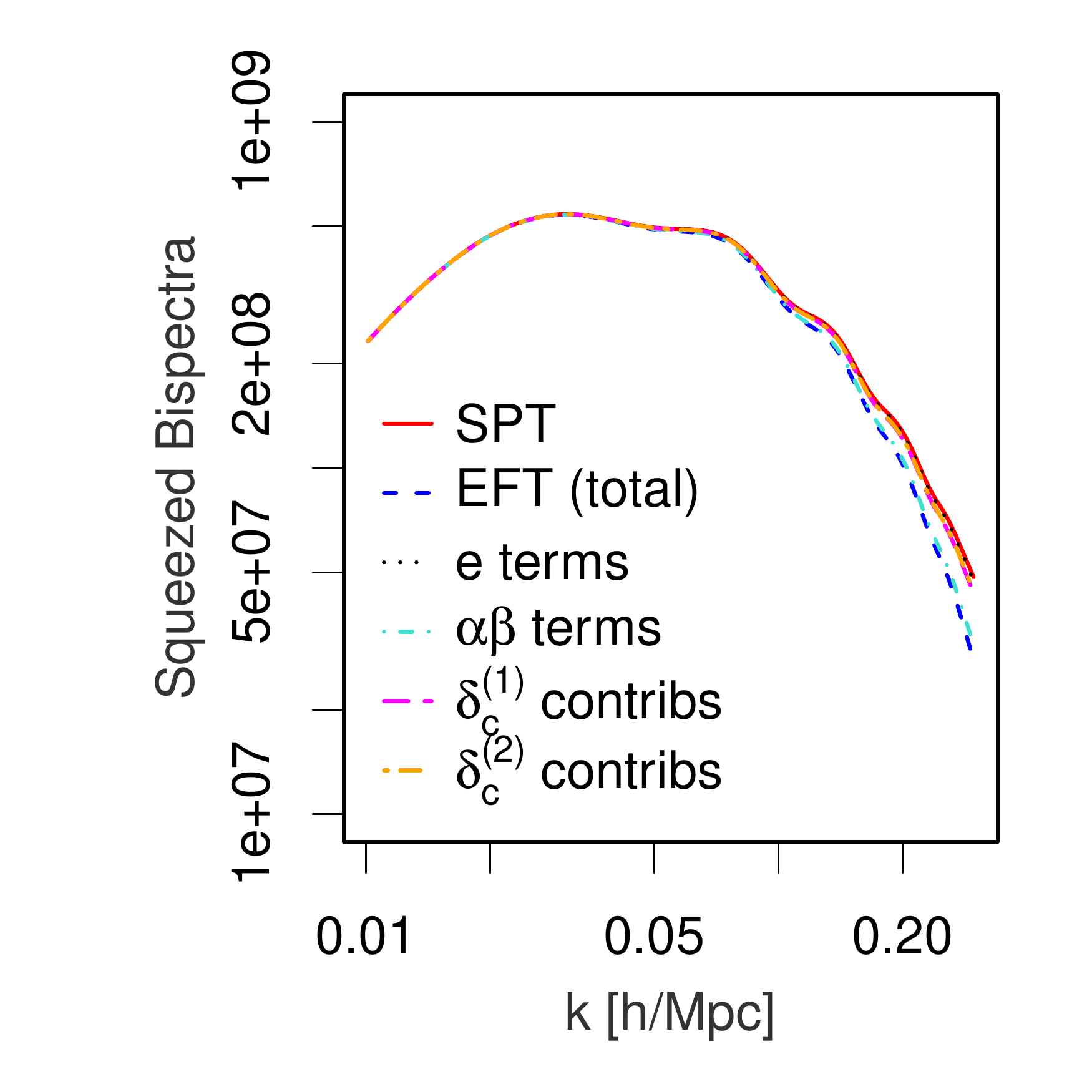}
\includegraphics[width=0.45\textwidth]{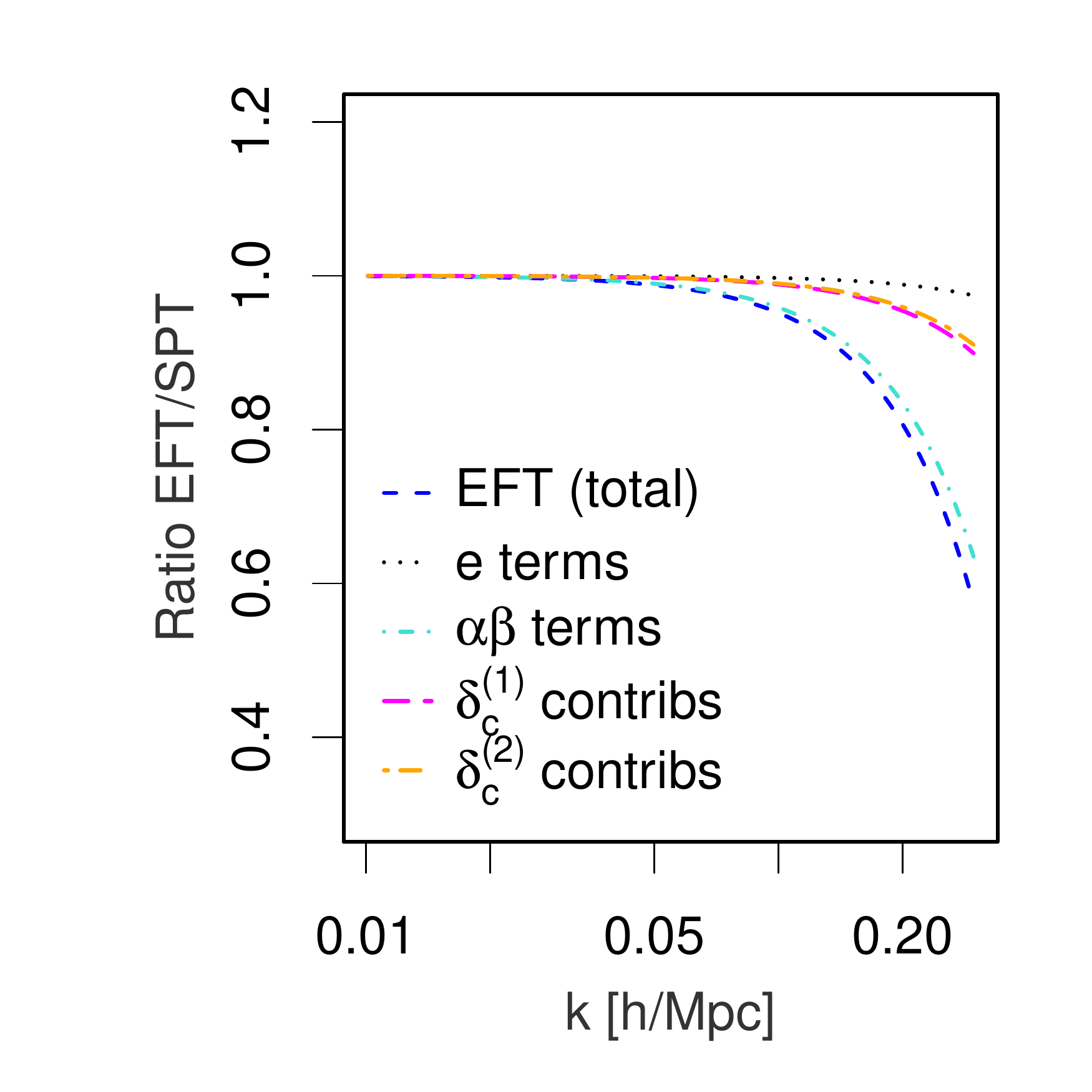}
\caption{(Left panel) Contributions to the squeezed matter bispectrum $B_\delta(\bk,-\bk,-\bq_b)$ from the various terms in Eq.~\eqref{eq:sq_EFTDelta} are depicted.
(Right panel) Ratio of various contributions to EFT and the SPT are shown. The computation of the squeezed bispectrum is peformed using the same set up
as in Fig.\ref{fig:bSPT_bEFT}}
\label{fig:bSPT_contrib}
\vspace*{2em}
\end{figure}

\para{Parametrization of the squeezed limit}
We employ the following parametrization for the bispectrum to analyse its squeezed limit:
\ben
&& {\bf k}_1 = {\bf k} - {\bf q}_a;\quad {\bf k}_2 = -{\bf k} + {\bf q}_a+{\bf q}_b;\quad {\bf k}_3 = -{\bf q}_b\,,
\label{eq:triangle}
\een
and take the limit $q_a,q_b\rightarrow 0$. We denote the cosine of angle between ${\bk}$ and ${\bq}_a$ (${\bq}_b$) as $\mu_{ak}$ ($\mu_{bk}$), and that between ${\bq}_a$ and $\bq_b$ as $\mu_{ab}$.
Expanding the wavenumbers in a Taylor series about $k$, and keeping only the leading powers of ${q_a/k}$ and ${q_b/k}$, gives 
\ben
&& k_1 = k\left ( 1- \mu_{ak}{q_a \over k}\right )\,, \quad
k_2 = k\left ( 1- \mu_{ak}{q_a \over k}-\mu_{bk}{q_b \over k} \right )\,, \quad 
k_3 = q_b\,.
\een
The power spectra are similarly expanded in the form:
\ben
&& P(k_1) = P(k)\left [1 -{q_a\mu_{ak} \over k}{d\ln P(k) \over d\ln k}\right ]\,, \;\;\nn\\
&& P(k_2)=P(k)\left [1 -{1 \over k}({q_a\mu_{ak} +q_b\mu_{bk}}){d\ln P(k) \over d\ln k}\right ]\,,\;\; P(k_3)=P(q_b)\,.
\een
To simplify notation we shall, in addition, assume a locally power-law power spectrum with $P(k)\propto k^n$ throughout. 

\para{Taking the squeezed limit of the density bispectrum} Taking the limit $q_a, q_b\rightarrow 0$, one finds that the results retain a dependency on the angle $\mu_{bk}$. We will take the angle-averaged value of this quantity, such that, for example, $\langle \mu_{bk}\rangle =0$, and $\langle \mu_{bk}^2\rangle =1/3$. The formulae for the SPT and EFT-only contributions to the density bispectrum are given in Eq.~\eqref{eq:SPT_tree} and Eq.~\eqref{eq:B_EFT_contribs}, respectively.
The SPT part of the bispectrum $B_\delta^{\rm SPT}$  has the following contribution:
\ben
&& \lim_{\bq_a,\bq_b\rightarrow 0} B^{\rm SPT}_{\delta}(\bk-\bq_a,-\bk+\bq_a+\bq_b, -\bq_b)  \\
&&\hspace{15mm}  \neweq\left [{13\over 7} +{8 \over 7}\left ( {\bk\cdot \bq_b \over k q_b}\right ) -\left ( {\bk \cdot \bq_b \over k q_b} \right )^2{d\ln P(k)\over d\ln k} \right ] P(k) P(k_3)
= \left [ {47 \over 21} -{n \over 3}\right ] P(k) P(k_3)\,.\nn
\label{eq:sq_EDS}
\een
The squeezed limit for the EFT contributions may be evaluated similarly, resulting in (where we recall  $B^{\text{EFT}}_{\delta}= B_{\delta_c^{(1)}}+B_{\delta_c^{(2)},\delta}+B_{\delta_c^{(2)},e}+B_{\delta_c^{(2)},\alpha\beta}$)
\bes\label{eq:sq_EFTDelta}
\ben
&& B_{\delta_c^{(1)}}\neweq -\xi \Big[{33-7 n\over 21} \Big] k^2 P(k) P(k_3)\,,\\
&& B_{\delta_c^{(2)},\delta}\neweq  -\xi M_d\Big[  { 61 - 7 n\over 21} \Big] k^2 P(k) P(k_3)\,,\\
&& B_{\delta_c^{(2)},e}\neweq  -\xi M_d\Big[ 4 {\epsilon_1 \over \xi} \Big] k^2 P(k) P(k_3)\\
&& B_{\delta_c^{(2)},\alpha\beta}\neweq  -\xi {4 \Big[ (87+66 m_d + 12 m_d^2) - (11+4 m_d) n \Big] \over 3 (2+m_d)(9+2 m_d)} k^2 P(k) P(k_3)\,.
\een
\ees
That each of the contributions appears as a coefficient times $k^2 P(k) P(k_3)$ is typical of the EFT terms, and may have been inferred by the form of the effective stress-energy tensor, $\tau_\theta$, in Eq.~\eqref{eq:ansatz_noPNG}.

In Fig.~\ref{fig:bSPT_bEFT} we plot the results for the squeezed limits of the SPT (red, solid) and total EFT (blue, dashed) density bispectra utilising the coefficients given in Eq.~\eqref{eq:params}, and choosing the scaling coefficient as $m_d=1$. We assume a factor of $10$ ratio between the long and short modes. It is clear that EFT contributions become important to the squeezed limit already at $k\approx 0.05 h$/Mpc. We find that the $k$-dependent correction can contribute up to $40\%$ of the
total estimate at $k=0.25 {\rm hMpc}^{-1}$. This is in rough agreement 
with previous findings where a factor of two improvement was reported when EFT corrections are included \citep{bispec}. To demonstrate that our results are robust to the choice of scaling parameter, we have plotted in Fig.~\ref{fig:bSPT_approx} a comparison to the results obtained with $m_d=(1-n)/(3+n)$, as suggested by a self-similar solution within an EdS Universe. Furthermore, in Fig.~\ref{fig:bSPT_contrib} we plot the individual contributions from the various terms in Eq.~\eqref{eq:sq_EFTDelta}. It is apparent that the so-called `$\alpha\beta$' terms, originating from the cross-correlation between the SPT $\delta^{(n)}$ and counter-terms $\delta^{(n)}_c$, are the dominant contributors to the EFT bispectrum.

\subsubsection{Projected or 2D surveys}
\label{sec:2D}

\begin{figure}
\centering
\text{\hspace{1cm} Projected (2D) Bispectrum - $\delta$}\par\smallskip
\includegraphics[width=0.45\textwidth]{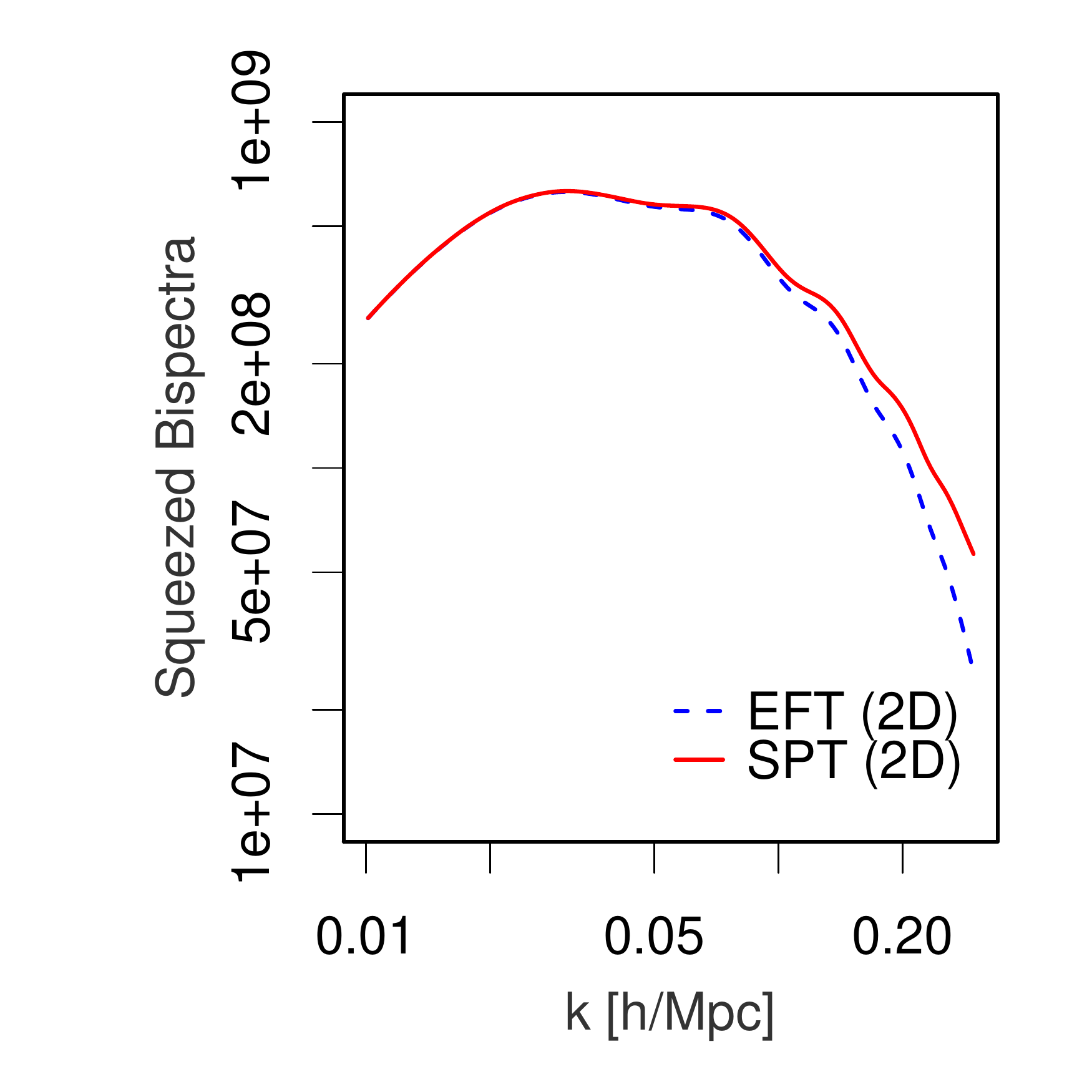}
\includegraphics[width=0.45\textwidth]{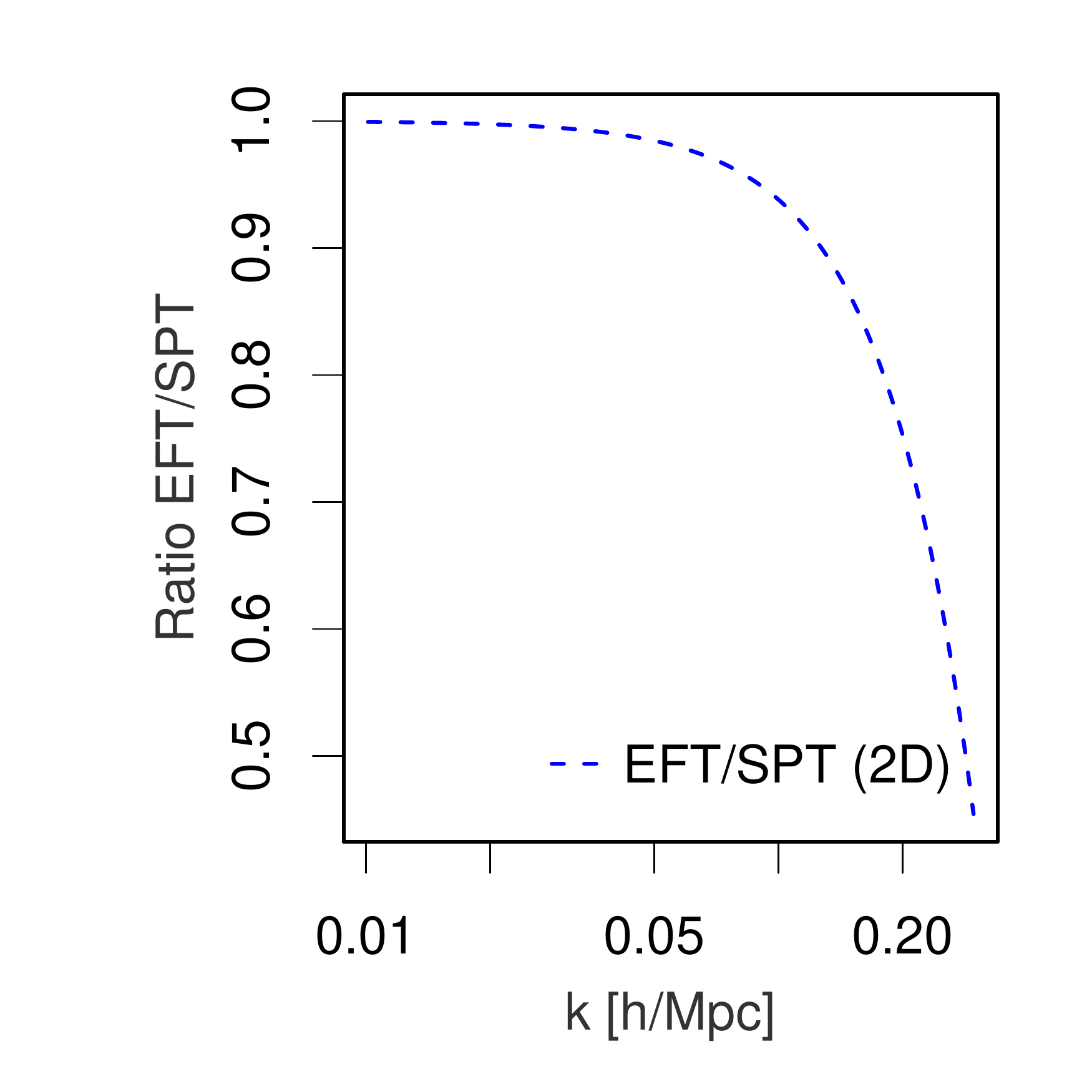}
\caption{As with Fig.~\ref{fig:bSPT_bEFT} but restricted to $2$D [ see Eq.(\ref{eq:proj1})-Eq.(\ref{eq:proj4}) ].
The parameters used for the 2D calculation are same as those used for 3D.}
\label{fig:bSPT_bEFT_2D}
\vspace*{2em}
\end{figure}

Before the advent of redshift surveys, initial studies of galaxy clustering were
performed using projected (or 2D) galaxy surveys. 
 Some of the earliest results in this directions were obtained using
the Lick Catalogue \citep{Lick}. One of the most well known
2D galaxy surveys the APM survey contained $1.3 \times 10^6$ galaxies and allowed 
the first clear studies of higher-order correlation functions \citep{Ber3}. Indeed while redshift surveys allow one to probe the
clustering in 3D, even the current generation of 3D surveys do not contain a much larger 
number of galaxies, though the situation is steadily improving.
The main limitations of 
2D surveys comes from mixing of 3D modes due to projection which are at a different
stages of non-linearities. Indeed many deprojection algorithms have also been
suggested \citep{George}. The redshift surveys of course in contrast suffer from mixing of
density and velocity field characteristics (see discussion in \textsection\ref{sec:RSD}). 

In order to compute the squeezed limits in the 2D case one utilises the same expressions as for the 3D case but
adjusts the average of the cosine angle to $\langle\mu^2_{bk} \rangle ={1/2}$. The SPT and EFT (c.f. Eq.~\eqref{eq:B_EFT_contribs}) squeezed limits are given by:
\bes
\ben
\label{eq:proj1}
&& B^{\text{SPT}}_{\text{2D}}\neweq\left [ {17 \over 7}-{n\over 2} \right ]P(k)P(k_3)\,,\\
&&B_{\delta_c^{(1)}}\neweq -\xi \frac{20-7 n}{14}k^2 P(k)P(k_3)\,,\nn\\
&&B_{\delta_c^{(2)},\delta}\neweq -\xi M_d \frac{48-7 n}{14} k^2 P(k)P(k_3)\,,\nn\\
&& B_{\delta_c^{(2)},e} \neweq -\xi M_d \left [4{\epsilon_1\over \xi}+{2\over 3}{\epsilon_2\over \xi}+ {1\over 3}{\epsilon_3\over \xi}\right ]k^2 P(k)P(k_3)\,,\nn\\
&& B_{\delta_c^{(2)},\alpha\beta} \neweq  -\xi\,
{2\left [(56+44 m_d+8 m_d^2) - (11 + 4 m_d) n\right ]\over (2+m_d)(9 + 2 m_d)} k^2 P(k)P(k_3)\,. \nn
\label{eq:proj4}
\een
\ees
In contrast to the 3D case the $\epsilon_2$ and $\epsilon_3$ terms  do not vanish, suggesting that 2D surveys may be useful to break degeneracies between the EFT contributions.
We have also used the Limber approximation $P(k)\approx P(k_{\perp})$ as $\bk=\bk_{\parallel}+\bk_{\perp}$ with $k_{\parallel}\ll k_{\perp}$. The SPT and (total) EFT contributions are plotted in Fig.~\ref{fig:bSPT_bEFT_2D}.
While the results above are derived for projected galaxy surveys, the formulae are equally valid
for, say, projected weak lensing surveys with an appropriate choice for the selection functions.

\subsection{Squeezed limit of the EFT velocity divergence bispectrum}\label{sec:VelDiv}
\begin{figure}
\centering
\text{\hspace{1cm} Divergence of Velocity in 3D - $\bar\theta$}\par\smallskip
\includegraphics[width=0.45\textwidth]{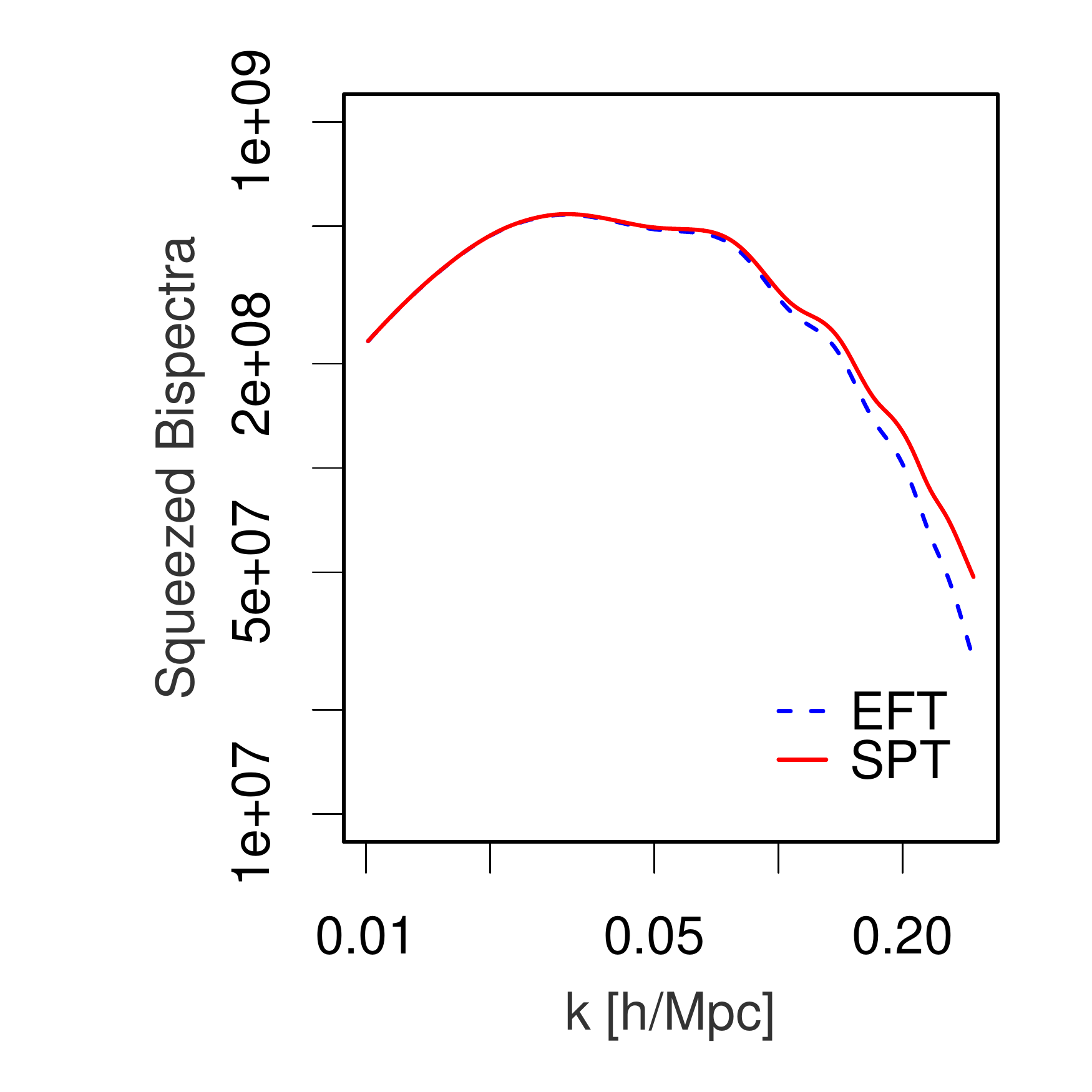}
\includegraphics[width=0.45\textwidth]{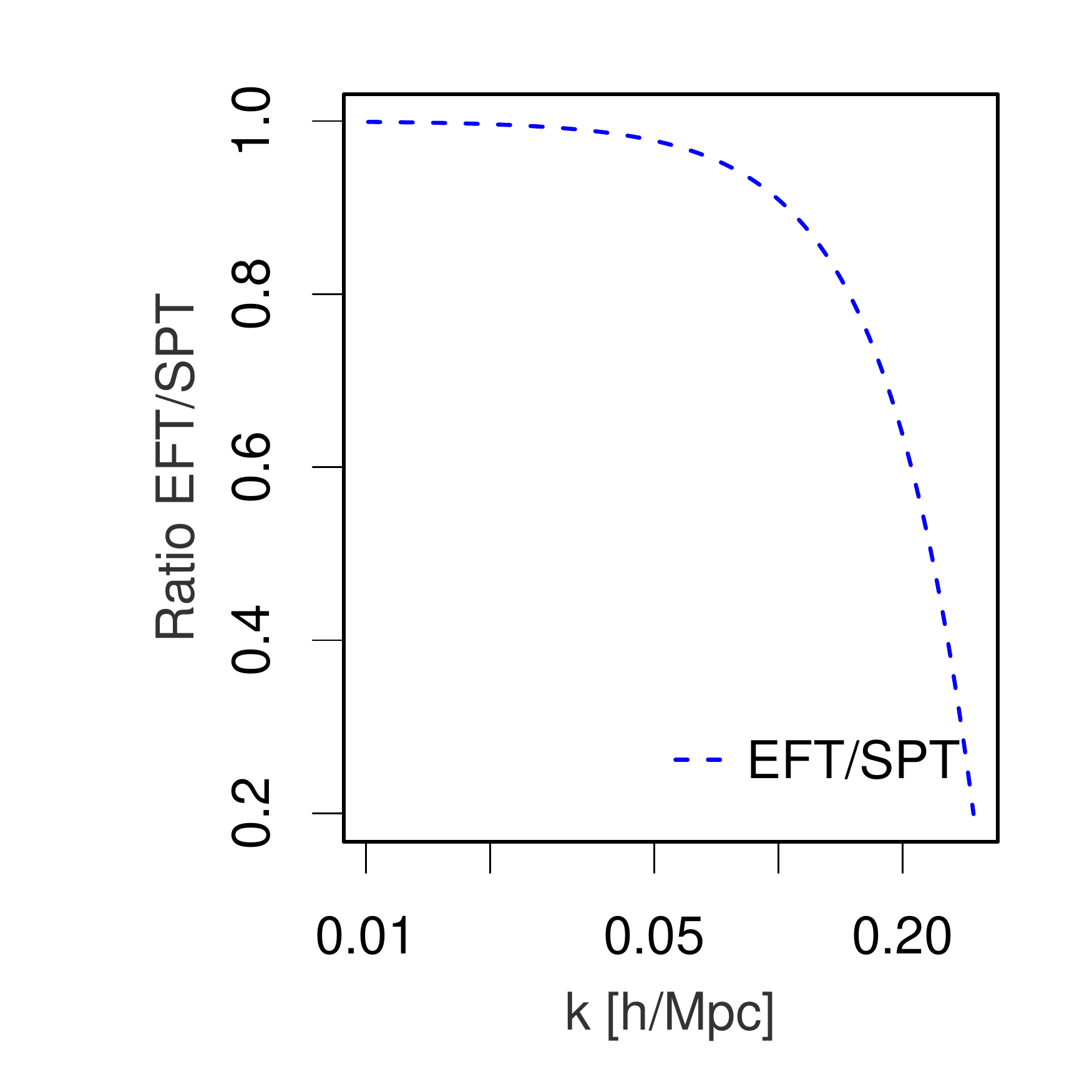}
\caption{(Left panel) Squeezed velocity divergence bispectrum $B_{\bar{\theta}}(\bk,-\bk,-\bq_b)$ for $q_b\equiv k_3=k/10$, for the both SPT (red, solid) and EFT (blue, dashed). (Right panel) Ratio of the EFT and SPT squeezed bispectra. The expressions for various contributions are presented in Eq.(\ref{eq:theta_SPT_sq})-Eq.(\ref{eq:sq_EFTTheta}).}
\label{fig:bSPT_bEFT_theta}
\vspace*{2em}
\end{figure}

Extension of the results for the squeezed limit of the density bispectrum given in Sec.~\ref{sec:Density} to the velocity divergence may be easily inferred using the results of Sec.~\ref{sec:eftoflss}. In particular, the SPT contribution of Eq.~\eqref{eq:SPT_Bisp2} gives
\ben\label{eq:theta_SPT_sq}
B_{\bar \theta}^{\text{SPT}}\neweq \left [ {31\over 21}-{n\over 3}\right ]P(k)P(k_3)\,,
\een
while the EFT-\textit{only} contributions given in Eq.~\eqref{eq:veldiv_EFT} result in the squeezed limits 
\bes\label{eq:sq_EFTTheta}
\ben
&& B_{{\bar \theta}_c^{(1)}}\neweq -\xi (2+m_d)\Big[{17-7 n\over 21} \Big] k^2 P(k) P(k_3)\,,\\
&& B_{{\bar \theta}_c^{(2)},\delta}\neweq  -\xi (3+m_d) M_d\Big[  { 61 - 7 n\over 21} \Big] k^2 P(k) P(k_3)\,,\\
&& B_{{\bar \theta}_c^{(2)},e}\neweq  -\xi (3+m_d) M_d\Big[ 4 {\epsilon_1 \over \xi} \Big] k^2 P(k) P(k_3)\\
&& B_{{\bar \theta}_c^{(2)},\alpha\beta}\neweq  -\xi {4 \Big[ (27+8 m_d + 2 m_d^2) - (15+10 m_d+2 m_d^2) n \Big] \over 3 (2+m_d)(9+2 m_d)} k^2 P(k) P(k_3)\,.
\een
\ees
Using the same choice of parameters as for Fig.~\ref{fig:bSPT_bEFT}, we plot, in Fig.~\ref{fig:bSPT_bEFT_theta}, the SPT and EFT predictions for the squeezed limit of the velocity divergence bispectrum. Again it is apparent that EFT contributions can be significant for $k\gtrsim 0.05 h/$Mpc. This is significant in that divergence of the peculiar velocity, $\theta$, is known to be an important probe of $\Omega_{\rm M}$ \citep{Ber2} and thus of theories of modified gravity.

\subsection{Squeezed limit of the EFT bispectra of Primordial Non-Gaussianity}\label{sec:squ_PNG}
\begin{figure}
\centering
\text{\hspace{1cm}Primordial Non-Gaussianity}\par\smallskip
\includegraphics[width=0.45\textwidth]{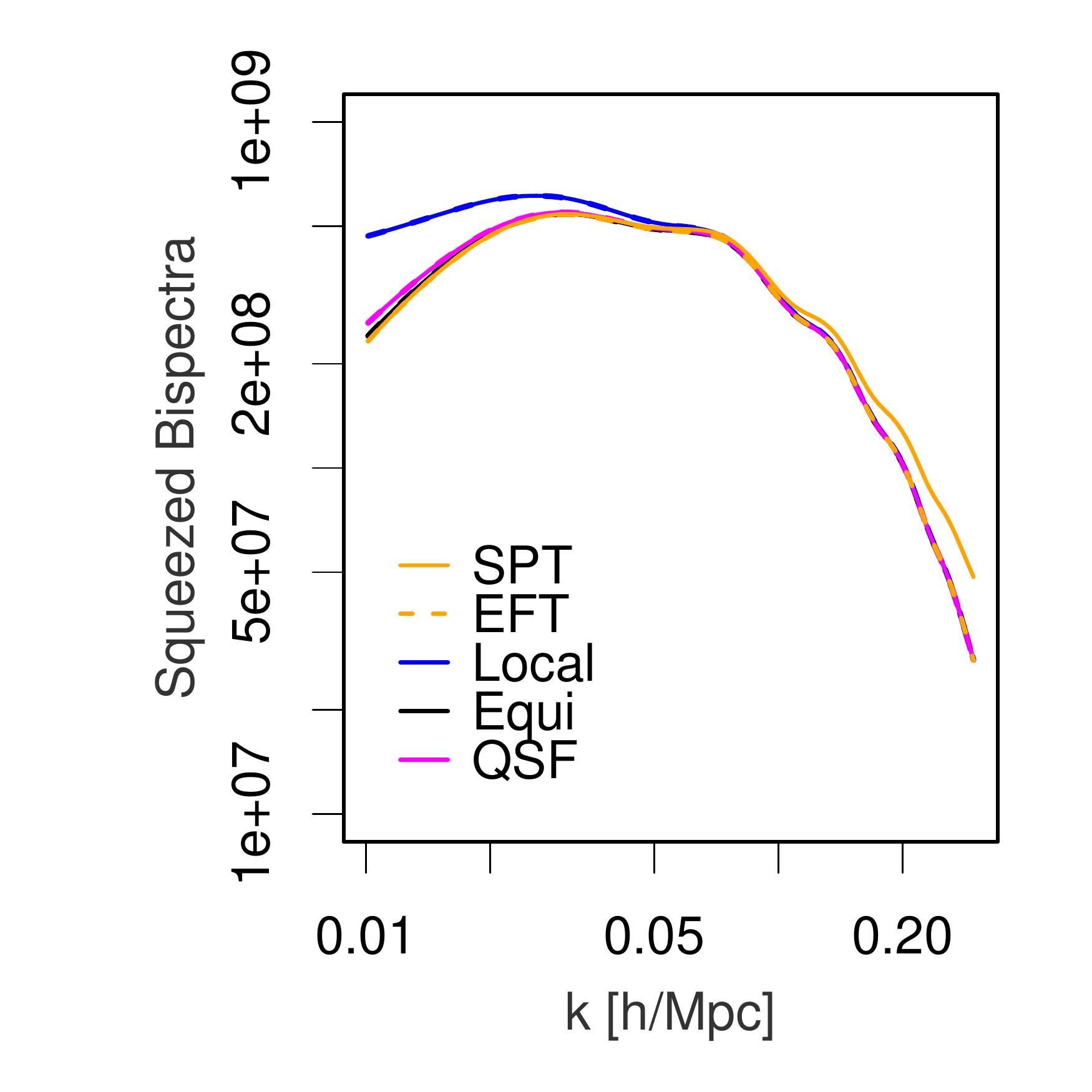}
\includegraphics[width=0.45\textwidth]{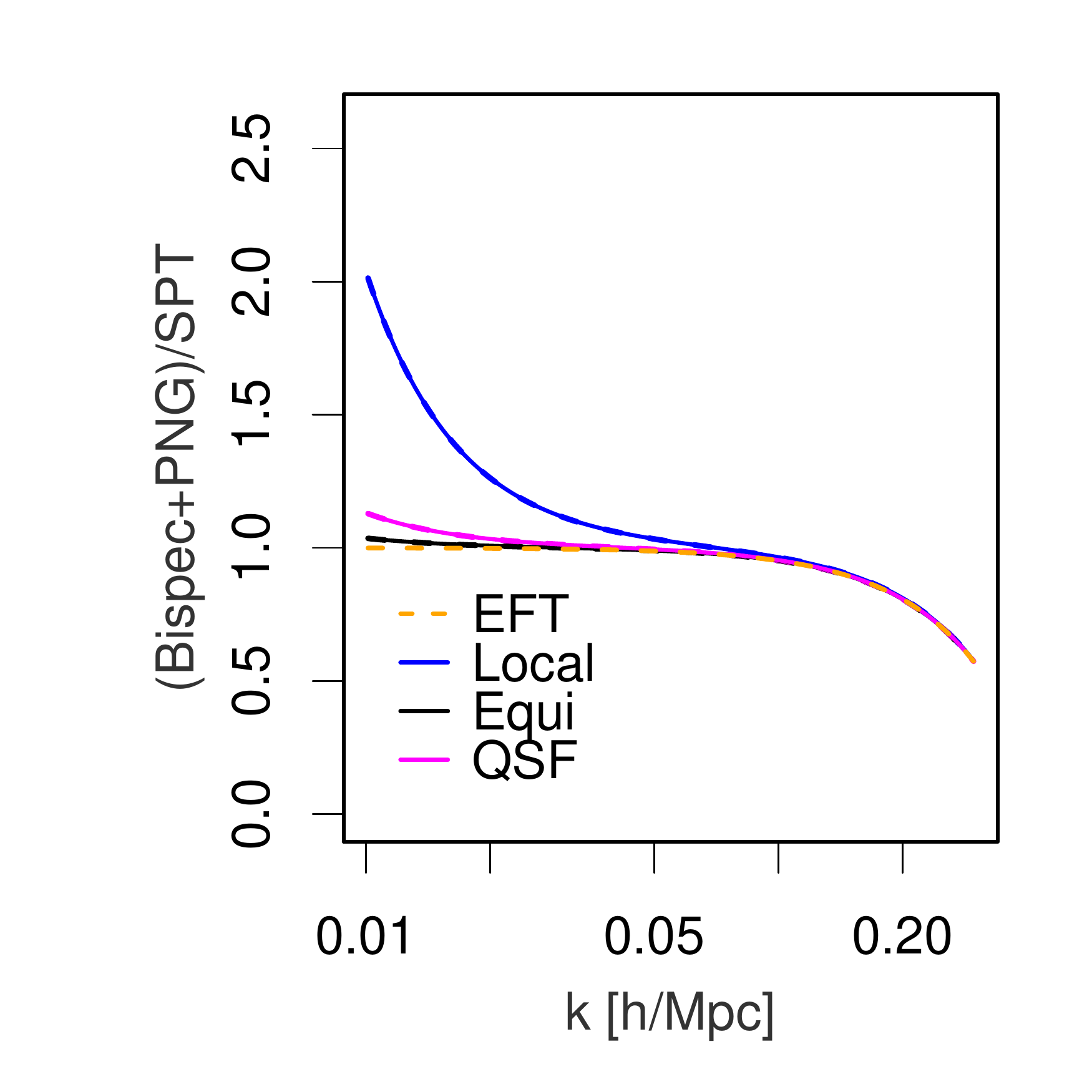}
\caption{(Left panel) Squeezed bispectra $B_{\delta}(\bk,-\bk,-\bq_b)$ for $q_b\equiv k_3=k/10$ for SPT (solid) and EFT (dashed). 
The cases of zero primordial non-Gaussianity $f_{\rm NL}=0$ is plotted in orange, while the local, equilateral and quasi-single field
bispectra (added to the respective SPT or EFT gravitational bispectrum) with $f_{\rm NL}=10$ are plotted in blue, black and magenta, 
respectively. The models are defined in Eq.(\ref{eq:local})-Eq.(\ref{eq:qsf}). We have taken the following 
parameter values $m_g$=1 and ${\gamma} = {\gamma}_i =1$.
It is clear that the EFT corrections to the inflationary bispectrum contribution are negligible compared to the effect 
on the gravitational bispectrum. The expressions from various terms are listed in Eq.(\ref{eq:sq_PNG_EFT_only}).
(Right panel) Ratio of the various bispectra shown in the left panel to the SPT-\textit{only} contribution.}
\label{fig:bSPT_bEFT_PNG}
\vspace*{2em}
\end{figure}

In Sec.~\ref{sec:PNG} we described how primordial non-Gaussianity (PNG) may be incorporated into the EFT framework. Our prediction for the additional contributions to the bispectrum were listed in Eq.~\eqref{eq:PNG_SPT_EFT}, with the bispectra for the particular primordial models (\textit{local, equilateral and quasi-single field}) being considered for this paper listed in Eq.~\eqref{eq:PNG_models}. The squeezed limits of the EFT-\textit{only} contributions are then given (where we recall our assumption that $m_g$ is the scaling dimension for all of the EFT counterterms for PNG)  in the angle-averaged case by 
\bes\label{eq:sq_PNG_EFT_only}
\ben
&& B_{\text{PNG}}^{g(1)}\neweq 2 \gamma(a) k^2 B_{\text{PNG}}^{\text{SPT}}(k,k,k_3)\,,\\
&& B_{\text{PNG}}^{g(2)}\neweq \gamma(a){  m_g (5 +2 m_g) \over 3 (1+m_g) (7+2 m_g)} \left( 7+ n\right)  k^2  P(k) P(k_3)    \,,\\
&& B_{\text{PNG}}^{g_1}\neweq 2\gamma_1(a) k^2 P(k) P(k_3) \,,\\
&& B_{\text{PNG}}^{g_2}\neweq -{4\over 3}\gamma_2(a) k^2 P(k)^2\,,\\
&& B_{\text{PNG}}^{\alpha\beta}\neweq {4\over 3(1+m_g) (7 + 2 m_g)}  \left(33+42 m_g + 12 m_g^2 - (7+4 m_g) n\right) P(k) P(k_3) \,.
\een
\ees
In Fig.-\ref{fig:bSPT_bEFT_PNG} (Left panel) we show the squeezed bispectra $B_{\delta}(\bk,-\bk,-\bq_b)$ for $q_b\equiv k_3=k/10$ for SPT (solid) and EFT (dashed) for the local, equilateral and quasi-single field models. 
The cases of zero primordial non-Gaussianity are plotted in orange, while the local, equilateral and quasi-single field bispectra 
(added to the respective SPT or EFT gravitational bispectrum) with $f_{\rm NL}=10$ are plotted in blue, black and magenta, respectively. We have chosen for simplicity that $\gamma, \gamma_1, \gamma_2$ take values of $1 h/\text{Mpc}$ and $m_g$ takes a value of unity. However, it is clear that the EFT corrections to the inflationary bispectrum contribution are negligible compared to the effect on the gravitational bispectrum, so the choices are relatively unimportant. This agrees with the results of \citep{primordial,Yvette}.
The ratio of the various bispectra -- individually shown in the left panel -- to the SPT-only contribution is plotted in the right panel, emphasising that the key distinguishing region for primordial non-Gaussianity occurs on large scales; this is most significant for the local model (scaling dimension $\Delta =2$), with the quasi-single field model showing mild uplift ($\Delta=1$), while distinguishing the equilateral model ($\Delta=0$) from the gravitationally induced non-Gaussianity appears difficult. These results, which demonstrate that EFT corrections to the primordial non-Gaussian bispectra are negligible, while the measurement of PNG using large scale structure requires large scale observations, are consistent with results in the literature \citep{Yvette,Tellarini:2015faa}. 

\subsection{Squeezed limit of the RSD bispectrum} \label{sec:squ_RSD}
In Sec.~\ref{sec:RSD} we recapitulated a discussion of the redshift-space distortion bispectrum in SPT from \citep{Thesis} and demonstrated the extension to the effective field theory framework. In this section we compute the squeezed limit
of the various contributions to the bispectrum.

\para{Parametrization of the squeezed limit} As the distortion occurs along the radial direction, additional angular
dependencies become important, as listed in Eq.~\eqref{eq:RSD_angles}. One can express the angular variables $\mu_i\equiv \hat{\bk}_i.\hat{\bx}_{\parallel}$ in terms of $\mu_k\equiv \hat{\bk}.\hat{\bx}_{\parallel}$ and the cosines of the angles between ${\bk}$, $\bq_a$ and $\bq_b$ described in Sec.~\ref{sec:Density} as, 
\ben
&& \mu_1 = \mu_k + {1 \over k}\left (q_a \mu_{ak}\mu_k -q_a\mu_a\right )+\cdots\,,\;\;\\
&& \mu_2 = -\mu_k + {1 \over k}\left (q_a\mu_k+q_b\mu_{bk} -q_a\mu_a\mu_k -q_b\mu_b\mu_k \right )+\cdots\,,  \\
&& \mu_3 = -\mu_b\,.
\een
Our squeezed limit involves again taking the limit $q_a, q_b\rightarrow 0$ and then performing the angle averaging with respect to $\hat{\bk}$ and $\hat{\bq}_b$ (in practice there is no dependence on $\hat{\bq}_a$ in the limit). 

\para{Taking the squeezed limit of the RSD bispectrum} First we consider the $\text{SQ}1$ and $\text{SQ}2$ terms listed in Eqs.~\eqref{eq:LOS1} and~\eqref{eq:LOS2} . The SPT contributions -- where we recall the parametrisation defined in (and below) Eq.~\eqref{eq:F2} -- give
\begin{subequations}
\ben
\label{eq:sq1A}
&& B_{\rm SQ1_1} \stackrel{\text{sq}}{\supseteq} \left [ {5 + 4\epsilon\over 3} - {n\over 3} \right ] P(k) P(k_3)\,,\\
&& B_{\rm SQ1_2}  \stackrel{\text{sq}}{\supseteq} \Big [{2\over 9} (5+4\epsilon - n)  \Big ]P(k) P(k_3)\,, \label{eq:sq1B}\\
&& B_{\rm SQ1_3}  \stackrel{\text{sq}}{\supseteq} \Big [{1\over 225} (39+28\epsilon-11 n)  \Big ]P(k) P(k_3)\,, \label{eq:sq1C}\\
&& B_{\rm SQ2_1}  \stackrel{\text{sq}}{\supseteq}   \left [ {5 + 4\epsilon^\prime - n\over 9}\right ] P(k) P(k_3)\,,\label{eq:sq2A} \\
&& B_{\rm SQ2_2}  \stackrel{\text{sq}}{\supseteq} \left [  {2(65+44\epsilon^\prime -13 n)\over 225}\right] P(k) P(k_3)\,,\label{eq:sq2B}\\
&& B_{\rm SQ2_3}  \stackrel{\text{sq}}{\supseteq} \left [ {69+36\epsilon^\prime - 17 n\over 525}\right] P(k) P(k_3)\,,\label{eq:sq2C}
\een
\end{subequations}
As detailed in Sec.~\ref{sec:RSD}, only these contributions are affected by the EFT terms. Performing the computations similarly to the SPT terms, one obtains additional contributions from effective field theory to $B_{\rm SQ1}$ and $B_{\rm SQ2}$. These will be dominated by the $\alpha\beta$ terms as usual which gives 
\ben
&& B_{\rm SQ1_1}^{\text{EFT}} \stackrel{\text{sq}}{\supseteq} \xi(a)\Big[ 2{87 - 11 n + (66 - 4n) m_d + 12 m_d^2 \over 3 (2+m_d)(9+2m_d)} \Big] k^2 P(k) P(k_3) \,,\nn\\
&& B_{\rm SQ1_2}^{\text{EFT}} \stackrel{\text{sq}}{\supseteq}  \xi(a)\Big[ 4{ 87 - 11n +(66-4n) m_d + 12 m_d^2   \over 9 (2+m_d) (9+2 m_d)}\Big] k^2 P(k) P(k_3)\,,\nn\\
&& B_{\rm SQ1_3}^{\text{EFT}} \stackrel{\text{sq}}{\supseteq}  \xi(a)\Big[2{ 621 - 121n +(518-44n) m_d + 100 m_d^2     \over 225 (2+m_d) (9+2 m_d)} \Big] k^2 P(k) P(k_3)\,,
\een
and
\ben
&& B_{\rm SQ2_1}^{\text{EFT}} \stackrel{\text{sq}}{\supseteq} \xi(a)\Big[4{3(-9+5n)+2(-4+5 n)m_d + 2(1+n) m_d^2 \over 9 (2+m_d)(9+2m_d)} \Big] k^2 P(k) P(k_3) \,,\nn\\
&& B_{\rm SQ2_2}^{\text{EFT}} \stackrel{\text{sq}}{\supseteq} \xi(a)\Big[8{ -399+195n+10(-16+13n)m_d + 2(5+13n)m_d^2 \over 225 (2+m_d)(9+2m_d)} \Big] k^2 P(k) P(k_3) \,,\nn\\
&& B_{\rm SQ2_3}^{\text{EFT}} \stackrel{\text{sq}}{\supseteq} \xi(a)\Big[4{-411+255 n + 10(-20+17 n)m_d + 2(1+17 n) m_d^2 \over 525 (2+m_d)(9+2m_d)} \Big] k^2 P(k) P(k_3).
\een
The terms denoted $B_{\rm NLB}$ and $B_{\rm FOG}$ by contrast remain unchanged due to EFT terms at the order we are computing, and their squeezed limits are given by \cite{Thesis}
\ben\label{eq:sq_NLB}
&& B_{\rm NLB,1} \neweq 2 P(k) P(k_3)\,,\quad
 B_{\rm NLB,2} \neweq {4\over 3} P(k) P(k_3)\,,\quad
 B_{\rm NLB,3} \neweq {2\over 9} P(k) P(k_3)\,,\nn\\
\een
and
\ben\label{eq:sq_FOG}
&& B_{\rm FOG,1} \neweq \left[1-{n\over 9}\right] P(k) P(k_3)\,,\hspace{15mm}
 B_{\rm FOG,2} \neweq \left[{22\over 45}-{2 n\over 15} \right] P(k) P(k_3)\,,\nn\\
 &&B_{\rm FOG,3} \neweq \left[{3\over 5}-{n\over 15}\right] P(k) P(k_3)\,,\hspace{12mm}
 B_{\rm FOG,4} \neweq  \left[{22\over 75}-{2 n\over 25}\right] P(k) P(k_3)\,,\nn\\
&& B_{\rm FOG,5} \neweq  \left[{13\over 105}-{n\over 21}\right] P(k) P(k_3)\,,\qquad 
 B_{\rm FOG,6} \neweq  \left[{13\over 175} - {n\over 35} \right] P(k) P(k_3)\,.
\een
These results for the squeezed RSD bispectrum should prove useful for upcoming surveys such as
EUCLID  or LSST who offer the possibility of more exact measurements of redshift-space
distortions. As for the velocity divergence, a key source of interest in these measurements will be the
possibility of probing the signature of modified gravity theories.

%
%

\section{Conclusions and Future Prospects}
\label{sec:conclu}
%
Many alternatives methods have been developed in recent years to tackle the problem of
complete characterization and estimation of gravity induced (secondary) and primordial
non-Gaussianity at the level of bispectrum; these include estimators such as the skew-spectrum \citep{MuHe}, the line correlation function \citep{Wolstenhulme:2014cla}, 
modal decomposition methods \citep{Modal,Regan:2011zq,Byun:2017fkz} and the integrated bispectrum or the position-dependent
power spectrum studied in this paper. The integrated 
bispectrum probes the three point correlation function in the squeezed limit and is linked to the well-known consistency
relations.

While standard perturbation theory has been demonstrated to work well on large scales, and high redshifts, in order to exploit the scope of upcoming data sets, it is necessary to incorporate the backreaction effects of UV physics. Typically regularized perturbation theory 
is employed to improve the performance of standard perturbation theory, while semi-analytical approaches such as the
halo model have also been used to extend the reach of predictions without requiring expensive N-body simulations. In this paper we have 
used the EFT approach which incorporates the effects of small scale physics by modelling their impact by symmetry considerations to renormalise the density and velocity divergence, via an effective stress-energy tensor, with parameters which are typically computed using small-scale numerical simulations.

Previous studies of the consistency relations (or equivalently the angle averaged squeezed bispectrum) have been performed using standard perturbation theory. In this
paper we have extended the results using EFT. Our main results are given in Sec.~\ref{sec:consistencyRels}, where in Sec.~\ref{sec:Density} and Sec.~\ref{sec:VelDiv} we compute these statistics for the density and velocity divergence, for which the computation of the bispectrum is recapitulated in Sec.~\ref{sec:eftoflss}. The consistency relations for the density field in the case of a 2D projection survey are also presented in Sec.~\ref{sec:2D}. We find that EFT corrections to the consistency relations can become significant on relatively large scales $k\gtrsim 0.05 h/\text{Mpc}$.
Having described the computation of effective field theory corrections to the bispectrum induced by models of primordial non-Gaussianity in Sec.~\ref{sec:PNG}, we have generalized these relations to the case of primordial
non-Gaussianity in Sec.~\ref{sec:squ_PNG}. It is apparent that EFT corrections to the standard formulae make negligible difference to the squeezed bispectrum. In addition the squeezed limit of the RSD bispectrum (described in detail in Sec.~\ref{sec:RSD}) is described in Sec.~\ref{sec:squ_RSD}.

Our results can be used trivially to compute the position dependent power spectrum
or, equivalently, the integrated bispectrum. Such calculations can also be performed in 
real space where the position dependent two-point correlation functions and the related integrated 
three-point correlation function (three-point correlation function 
in the squeezed limit) can be used in real or redshift space.
It is notable that the position-dependent correlation function from the SDSS-III BOSS DR10 CMASS sample has been measured at a significance of $7.4\sigma$ \citep{Chiang:2015eza}, highlighting the potential to strongly probe the consistency relations investigated here as data sets continue to improve.

The methods described in this paper are complementary to the related cumulant correlators \citep{Munshi1}.
The lowest order (two-to-one) cumulant correlators have been used to probe the {\em hybrid} bispectrum 
involving the kinetic Sunyaev-Zeldovich (kSZ) effect and weak lensing in \citep{Dore}.
The same approach was used to cross-correlate Lyman-$\alpha$ flux and 
weak lensing convergence in \citep{VielDas}. The techniques elucidated here for computing
the integrated bispectrum and its real-space counterpart  can also be used to probe such cross-correlations (Munshi et al. 2017; in preparation).  
Finally, it should be noted that the redshift space distortion results presented here have been computed using a {\em flat-sky} approximation. For future surveys with large sky-coverage an expansion in spherical harmonics will be required to best extract the available cosmological information.


\section*{Acknowledgements}
\label{acknow}
DM acknowledges support from the Science and Technology
Facilities Council (grant numbers ST/L000652/1). The research leading to these results has received funding from the European Research Council under the European Union's Seventh Framework Programme (FP/2007--2013) and ERC Grant Agreement No. 308082.
DM \& DR also acknowledge useful discussions with members of the
University of Sussex cosmology group.

\bibliographystyle{JHEP}
\bibliography{paper}

\end{document}

%% file: local-macro.tex
\def\neweq{\stackrel{\text{sq}}{\approx}}

\def\be{\begin{equation}}
\def\ee{\end{equation}}
\def\ben{\begin{eqnarray}}
\def\een{\end{eqnarray}}
\def\bes{\begin{subequations}}
\def\ees{\end{subequations}}
\def\nn{\nonumber}

\def\bk{{\bf k}}

\def\br{{\bf r}}

\def\bk{{\bf k}}
\def\bp{{\bf p}}

\def\bx{{\bf x}}
\def\2p{{(2\pi)^2}}

\def\be{\begin{equation}}
\def\ee{\end{equation}}
\def\beq{\begin{equation}}
\def\eeq{\end{equation}}
\def\ben{\begin{eqnarray}}
\def\een{\end{eqnarray}}

\def\nn{{\nonumber}}

\newcommand{\beqa}{\begin{eqnarray}}
\newcommand{\eeqa}{\end{eqnarray}}

\def\ikap0{{\cal J}_{\theta_0}(r)}

\def\one1{\langle \kappa_{(i)}\kappa_{(j)} \rangle}
\def\one{{[\bar \xi^{(ij)}]}}

\def\ba{\begin{eqnarray}}
\def\ea{\end{eqnarray}}

\def\bk{{\bf k}}
\def\bq{{\bf q}}

\def\fNL{f_{\text{NL}}}

\newcommand{\semibold}[1]{{\fontseries{b}\selectfont{#1}}}
\newcommand{\para}[1]{\par\vspace{2mm}\noindent\semibold{{#1.}---}\ignorespaces}